\pdfoutput=1
\documentclass[aps,prd,superscriptaddress,showkeys,onecolumn,nofootinbib,10pt]{revtex4-2}
\usepackage{preamble}
\begin{document}

\title{Einstein-Cartan pseudoscalaron inflation, reheating and nonthermal leptogenesis}

\author{Carlo Di Benedetto\orcidD{}}
\email{dibenede@roma2.infn.it}
\affiliation{University of Rome ``Tor Vergata'', via della Ricerca Scientifica 1, 00133 Roma, Italy}
\affiliation{INFN, Sezione di Roma ``Tor Vergata'', via della Ricerca Scientifica 1, 00133 Roma, Italy}

\author{Alessandro Di Marco\orcidA{}}
\email{alessandro.dimarco1@inaf.it}
\affiliation{Istituto Nazionale di Astrofisica,\\
Istituto di Astrofisica e Planetologia Spaziali (INAF-IAPS),\\
Via Fosso del Cavaliere, 100, 00133 Rome, Italy}
\affiliation{INFN, Sezione di Roma ``Tor Vergata'', via della Ricerca Scientifica 1, 00133 Roma, Italy}

\author{Emanuele Orazi\orcidB{}}
\email{orazi.emanuele@gmail.com}
\affiliation{School of Science and Technology and Physics Department, Federal University of Rio Grande do Norte, Campus Universit\'ario- Lagoa Nova, Natal-RN 59078-900, Brazil}

\author{Gianfranco Pradisi\orcidC{}}
\email{gianfranco.pradisi@roma2.infn.it}
\affiliation{University of Rome ``Tor Vergata'', via della Ricerca Scientifica 1, 00133 Roma, Italy}
\affiliation{INFN, Sezione di Roma ``Tor Vergata'', via della Ricerca Scientifica 1, 00133 Roma, Italy}

\date{\today}

\begin{abstract}
We study the postinflationary dynamics of an Einstein-Cartan-Holst gravity-motivated inflationary scenario, known as Einstein-Cartan 
pseudoscalaron inflation, coupled to a type-I seesaw extension of the Standard Model with three heavy right-handed Majorana neutrinos. In particular, we show that nonthermal leptogenesis emerges as a necessary and self-consistent mechanism for generating the observed baryon asymmetry of the Universe, mainly because of the universal coupling of the inflaton to the additional heavy Majorana fermions. The resulting framework provides theoretical predictions that are fully compatible with the latest cosmological constraints from the Cosmic Microwave Background, Baryon Acoustic Oscillations, and Big Bang Nucleosynthesis, as well as with neutrino oscillation experiments, for a wide range of the fundamental Barbero-Immirzi model parameter $\gamma$, which controls the inflationary and postinflationary phases. 
In particular, for $\gamma \sim -1/100$ and a lightest Majorana-neutrino mass $\gtrsim 10^{13}$ GeV, we find a scalar spectral index $n_s \sim 
0.970$, a tensor-to-scalar ratio $r \sim 0.004$, for a number of $e$-folds before the end of inflation $N_e \lesssim 60$, 
and a baryon-to-entropy ratio $n_B/s \sim 8.7 \times 10^{-11}$.

\end{abstract}


\maketitle
\tableofcontents

\section{Introduction}
\label{Introduction}

The cosmological inflation \cite{Starobinsky:1980te,Guth:1980zm,Linde:1981mu,Albrecht:1982wi,Hawking:1981fz,Linde:1983gd} (for reviews, see \cite{Linde:1990flp,Linde:2007fr,Olive:1989nu,Baumann:2009ds,Uzan:2015pfm}) provides a compelling framework
for addressing the main shortcomings
of the standard Hot Big Bang (HBB) cosmology and for 
explaining the generation of both primordial scalar perturbations -- responsible for 
the formation of the large scale structures and the presence of the (primary) temperature fluctuations
in the cosmic microwave background radiation (CMB) --
and of a hypothetical stochastic background of gravitational waves \cite{Mukhanov:1990me,Riotto:2002yw,Guzzetti:2016mkm}.

The simplest and most widely studied inflationary cosmology is the 
single-field slow-roll scenario, where inflation is 
driven by a (pseudo)scalar field $\phi$ (the inflaton) 
slowly evolving along a quasi-flat potential $V(\phi)$ \cite{Steinhardt:1984jj,Liddle:1994dx}.
As inflation ends, the inflaton rolls down its potential 
and begins to oscillate about the \emph{true vacuum}, 
decaying into Standard Model (SM) or Beyond-the-Standard-Model (BSM) particles 
and thereby reheating the Universe, 
leading to the formation of a high-temperature relativistic plasma 
that eventually marks the onset of the radiation-dominated epoch of Big Bang cosmology.
The \emph{reheating} phase (see \cite{Albrecht:1982mp,Abbott:1982hn,Turner:1983he,Shtanov:1993es} for pioneering works 
and \cite{Bassett:2005xm,Frolov:2010sz,Allahverdi:2010xz,Amin:2014eta,Lozanov:2019jxc} for reviews on further developments)
represents a crucial stage in the history of the Universe
and can be regarded as a dynamical environment in which
a reasonable high-energy particle content (with related interactions)
can give rise to a wide variety of physical mechanisms
with potentially rich cosmological implications.
Such mechanisms may be triggered and eventually completed \emph{during} either the reheating process itself or \emph{after} reheating has concluded. 

The hypothetical postinflationary content of the Universe should play a fundamental role in addressing some of the major open problems of modern particle cosmology, unexplained so far within the standard HBB evolution based on General Relativity and SM.
Among these, the origin of the matter-antimatter asymmetry, \textit{i.e.} the baryon asymmetry of the Universe (BAU) 
stands out as one of the most challenging puzzles.
A straightforward way to address this problem is to introduce some postinflationary baryogenesis mechanisms satisfying the so called 
``Sakharov conditions'' \cite{Sakharov:1967dj} for the baryonic sector, namely: 
\textit{(i)} violation of the baryon number ($B$);
\textit{(ii)} violation of charge conjugation ($C$) and charge-parity ($CP$); 
\textit{(iii)} occurrence under out-of-equilibrium conditions.
The baryon asymmetry of the universe may originate via several pathways (see
\cite{Ignatiev:1978uf,Yoshimura:1978ex,Dimopoulos:1978kv,Yoshimura:1979gy,Weinberg:1979bt,Kolb:1979qa,Fry:1980ph,Fry:1980bc,Harvey:1981yk,Dolgov:1982th,Kolb:1983ni,Affleck:1984fy,Linde:1985gh,Shaposhnikov:1987tw,Cohen:1987vi,Ellis:1987rw}
for pioneering works on the subject and \cite{Dolgov:1991fr,Riotto:1998bt,Riotto:1999yt,Dine:2003ax,Cline:2006ts,Shaposhnikov:2009zzb,Canetti:2012zc}
for complete reviews.).
Among them, a particularly appealing possibility is provided by leptogenesis 
(see \cite{Fukugita:1986hr,Luty:1992un,Covi:1996wh,Plumacher:1996kc,Barbieri:1999ma,Giudice:2003jh} 
for classical works on standard leptogenesis scenarios, 
\cite{Lazarides:1990huy,Murayama:1992ua,Campbell:1992hd,Kumekawa:1994gx,Murayama:1993em,Giudice:1999fb,Asaka:1999yd,Asaka:1999jb,Zhang:2023oyo} for 
the nonthermal or inflationary case, 
while for reviews see, {\it{e.g.}},  \cite{Buchmuller:2005eh,Buchmuller:2004nz,Davidson:2008bu,Fong:2012buy,Blanchet:2012bk}.). 
In this case,
Sakharov conditions for the lepton sector allow for lepton asymmetry production, 
subsequently converted into (a sufficient amount of) baryon asymmetry 
via nonperturbative electroweak sphaleron processes \cite{Kuzmin:1985mm,Arnold:1987mh,Arnold:1987zg,Fukugita:1990gb}.
In several well-motivated realizations, leptogenesis can be naturally embedded within extensions of the SM.
For instance, the observation of SM neutrino oscillations implies a 
non-vanishing and small mass $m_{\nu_i}$ for the $i$th active neutrino, 
thereby providing the unique (up to now) established (although indirect) evidence of BSM physics. 
One of the most economical and theoretically interesting solutions to explain the tiny (active) neutrino masses is provided by the so-called (type I, II, or III) \textit{seesaw mechanisms}, where a certain set of fundamental fields is added to the SM content. 
Regardless of the specific realization, 
the seesaw framework generically 
introduces particles with masses at some scale $\Lambda$ much higher than the electroweak scale proportional 
to the Higgs vacuum expectation value (vev) $v\sim 246$ GeV. As a consequence, 
one obtains a suppressed light-neutrino mass scale of order $v^2/\Lambda$, 
much lighter than the remaining SM fermion masses.
If in the leptonic sector of a seesaw model Sakharov's conditions are satisfied, it is possible to provide an elegant mechanism for the generation of both neutrino masses and matter-antimatter asymmetry. 
These \emph{seesaw leptogenesis models} 
are thus expected to provide reliable predictions that simultaneously account for
the observed BAU 
\cite{Fields:2019pfx,Cooke:2017cwo,ParticleDataGroup:2024cfk}
and the measured neutrino properties
\cite{DESI:2024mwx,Allali:2024aiv,Esteban:2024eli,Capozzi:2025wyn,Capozzi:2025ovi,KATRIN:2024cdt}.

The simplest realization of post-inflationary lepton asymmetry production 
is provided by the widely studied \textit{type-I seesaw leptogenesis} framework, 
in which a certain number of heavy Majorana sterile right-handed neutrinos (RHNs) $N_k$, 
with large bare masses $M_{N_k}$, are added to the Standard Model. 
The RHN Yukawa couplings to the SM Higgs and lepton fields generate, after electroweak symmetry breaking, 
an additional Dirac mass matrix.
By diagonalization, the resulting mass eigenstates naturally reproduce the tiny masses of the active neutrinos, 
while the corresponding change-of-basis mixing matrix $U$ --
analogous to the Cabibbo-Kobayashi-Maskawa (CKM) matrix in QCD --  
and the Majorana masses, induce lepton number $L$ as well as $C$, and $CP$-violating interactions.  
Consequently, the expected out-of-equilibrium decays of the RHNs (at some high energy scale) 
in the early Universe can generate the required lepton asymmetry necessary for BAU. 

The \emph{minimal} type I seesaw model involves just two sterile Majorana RHNs which provide two corresponding massive SM active neutrinos and a massless one \cite{Minkowski:1977sc,Frampton:2002qc,Raidal:2002xf}. 
It has the advantage of requiring  a relatively small number (14) of free parameters. 
The most popular scenario, instead, 
involves three sterile Majorana RHNs that provide a mass for all active SM neutrinos \cite{Yanagida:1980xy,Glashow:1979nm,Gell-Mann:1979vob}.  
In this case, the mixing matrix $U$ is just a three-by-three matrix\footnote{By analogy to the case with three Dirac active neutrinos, this matrix is sometimes called the PMNS matrix 
\cite{Pontecorvo:1957qd,Pontecorvo:1957cp,Maki:1962mu}.} 
and the model involves 21 free parameters, eventually organized via the so-called Casas-Ibarra (CI) parameterization \cite{Casas:2001sr}.
However, type I seesaw leptogenesis with \emph{hierarchical RHNs} 
(\textit{i.e.} $M_{N_1} \ll M_{N_2}, M_{N_3}$) is typically driven only by the lightest $N_1$ whose mass is assumed to be slightly smaller than half the inflaton mass ($M_{N_1}\lesssim m_\phi/2$),  
while the other two states are expected to be too heavy ($M_{N_2},M_{N_3}\gg m_{\phi}$) to play an explicit, significant cosmological role.
In particular, the maximum (and absolute) amount of lepton asymmetry produced through the decay of the $M_{N_1}$ state
is controlled by the Davidson-Ibarra (DI) bound \cite{Davidson:2002qv} on the $CP$ violation parameter $\varepsilon_1$.
Data available on oscillations and solar neutrinos are not sufficient to fix all the details
of a (type I) seesaw model -- even adopting the CI recipe -- 
but it is always possible to check the presence of a window of model parameters,
largely consistent with experiments.

In general, two patterns of (type I seesaw) leptogenesis are possible,
depending on the \emph{nature} of the RHN $N_1$ production: \emph{thermal} and \emph{non-thermal}.  
In the thermal scenario, the inflaton field is not directly coupled to $N_1$ and decays exclusively into light SM degrees of freedom.
As a consequence, the corresponding hot and dense relativistic plasma can efficiently and \emph{thermally} produce the $N_1$ particles through the related SM interactions.
The subsequent generation of a lepton asymmetry depends on the model properties ($N_1$ mass, Yukawa couplings) 
and on the expansion rate of the Universe.
In general, two qualitatively different scenarios can be identified.
On the one hand, the $N_1$ particles may almost thermalize and then rapidly decouple from the thermal bath, allowing the lepton asymmetry to be generated efficiently on short cosmological timescales, possibly before the completion of reheating, at temperatures $T > T_{\rm reh}$.
On the other hand, $N_1$ particles can completely thermalize and remain coupled to the plasma 
for a prolonged period before going out-of-equilibrium and decaying efficiently, 
so that the lepton asymmetry is produced only on much longer cosmological timescales, 
after reheating has completed, at $T < T_{\rm reh}$.
The latter case just corresponds to the \textit{standard vanilla thermal leptogenesis} occurring during the early radiation-dominated phase 
and usually discussed in the literature.
In this vanilla case, natural 
bounds on the heavy neutrino mass (\textit{e.g.} $> 10^9 -10^{10}$ GeV) constrain the reheating temperature
(or the maximum temperature during reheating) to be very high.
This is typically problematic in supersymmetric extensions of the SM, where high temperatures lead to a copious production of gravitinos \cite{Weinberg:1982id,Ellis:1982yb,Nanopoulos:1983up,Ellis:1984eq,Khlopov:1984pf,Kawasaki:1994af,Giudice:2008gu,Kawasaki:2004qu,Kawasaki:2008qe,Kawasaki:2017bqm,Khlopov:2025pub},
which can potentially compromise the predictions of Big Bang Nucleosynthesis (BBN).

In the non-thermal scenario, on the other hand, the inflaton field directly couples to $N_1$, which is produced \emph{nonthermally} through inflaton decays.
Because of the typical large mass, the lightest  RHNs  $N_1$ are almost non-relativistic at production and 
hardly experience a (crucial) phase of thermal equilibrium, 
decaying into radiation and generating a lepton asymmetry before the reheating has completed. 
However, it is important to observe that, in some models, 
it could also happen that $N_1$ particles are driven (close) to thermal equilibrium for an extended period before their decay,
even \textit{after} reheating completion.
In such cases, the generation of the lepton asymmetry would become 
\emph{dynamically} equivalent to that of the standard vanilla thermal leptogenesis discussed before.
In nonthermal leptogenesis there is no direct connection between the reheating temperature and the mass of the right-handed neutrinos. 
This decouples the two scales, thereby avoiding a hypothetical gravitino problem.

This work investigates the post-inflationary reheating phase 
of a recently introduced (non-linear) Einstein-Cartan-Holst class of slow-roll inflationary models, 
also known as \emph{Einstein-Cartan pseudoscalaron inflation} \cite{Pradisi:2022nmh,DiMarco:2023ncs}, 
in which the fundamental (pseudo)scalar inflaton field arises from a dynamical component of the torsional degrees of freedom, which is absent in a purely Riemannian geometry. 
In order to get leptogenesis, the inflationary sector is coupled to a matter sector that includes the SM of particle physics  
extended to a type-I seesaw model by the addition of three hierarchical heavy sterile Majorana right-handed neutrinos.
A crucial observation is that, in these models, matter couplings to the inflaton originate from the minimal coupling to the torsional spin connection entering the covariant derivative. As a result, the inflaton couples in a \emph{universal} manner to the entire fermionic sector, SM particles 
and RHNs included. Moreover, the inflaton decay rates into fermions are proportional to the masses of the fermion themselves. Consequently, they are naturally vanishing for SM fermions in the unbroken Higgs phase and
very small, if compared to those of the inflaton to RHNs, also in the broken Higgs phase.
The mentioned peculiar properties lead to a particularly interesting reheating dynamics.
The key result is that a compulsory nonthermal leptogenesis mechanism driven by $N_1$ emerges. 
In particular, the inflaton decay amplitude to $N_1$ is relatively small, 
while $N_1$ exhibits sizable decay amplitudes to SM fermions.
Therefore, the inflaton field dominates the dynamics, with its lifetime that essentially determines both the duration of the reheating phase and the reheating temperature, 
whereas the relevant $N_1$s efficiently and rapidly generate the radiation plasma and the lepton asymmetry.
The global cosmological evolution depends on the specific details of the inflationary models, in particular on the value of the
fundamental Barbero-Immirzi parameter.
Notably, a broad range of values of the Barbero-Immirzi parameter produces robust predictions for both CMB observables 
and the final baryon asymmetry, 
in full agreement with current experimental data.

The paper is organized as follows.
Section 2 reviews the Einstein-Cartan-Holst gravity that gives rise to the single-field slow-roll inflationary scenario. 
It also reanalyzes  
the corresponding inflationary predictions, 
verifying their compatibility with the most recent observational data. Section 3 is devoted to the postinflationary stage. 
The analysis focuses on the properties of the model around the vacuum state and on the formulation of a reliable set of Einstein-Boltzmann equations jointly describing 
reheating and nonthermal leptogenesis. 
Numerical and analytical (approximate) solutions are presented and discussed.
Finally, in Section 4 the main results are summarized, and open directions are illustrated.
Some technical calculations can be found in the Appendices: Appendix A contains the adopted fermion conventions, 
while Appendix B reports structure and parameters of a general Boltzmann system describing nonthermal leptogenesis.

In this manuscript, \emph{natural units} $(\hbar = c = 1)$ are used, the reduced Planck mass is defined as $M_P = 1/\sqrt{8\pi G_N}$ with $G_N$ denoting Newton’s gravitational constant, 
and the four-dimensional spacetime signature is taken to be the mostly minus\footnote{Note that this convention is opposite to that used in \cite{Pradisi:2022nmh,DiMarco:2023ncs}.} $(+---)$.

\section{Einstein-Cartan pseudoscalaron inflationary models}
\label{Einstein-Cartan Pseudoscalaron Models}

In this Section, the Einstein-Cartan-Holst class of models, recently introduced in \cite{Pradisi:2022nmh,DiMarco:2023ncs}, is reviewed\footnote{See, also, \cite{Salvio:2022suk,Salvio:2025izr}}.
In an effective field theory approach aiming to describe the coupling of the Standard Model of particle physics to gravity, 
one can consider Einstein-Cartan theories (see \cite{Hehl:1976kj,Shapiro:2001rz,Hammond:2002rm} for some initial contributions and 
\cite{Choudhury:2014hja,Shaposhnikov:2020frq,Shaposhnikov:2020gts,Shaposhnikov:2020aen,Karananas:2021zkl,Desai:2015haa,Piani:2022gon,He:2023vlj} for recent progress and applications in particle physics and cosmology)
where the metric connection is promoted 
\'a la Palatini to be an independent field with respect to the metric or, better, to the vierbein.  Generically, the connection one-form can have 
torsion. Moreover, in an Einstein-Cartan spacetime the tangent bundle possesses locally flat bases, allowing the introduction of spinor fields in 
curved spacetime, a mandatory property to include Standard Model matter leptons and quarks besides fields with integer spin. Also gravitinos can be 
introduced, bringing to locally supersymmetric extensions ({\it{i.e.}} supergravities) that emerge in a natural way as effective theories of a more 
fundamental ultraviolet completion, like (super)string theory or M-theory. 

A generic Einstein-Cartan connection can always be obtained by summing a true tensor, the contortion, to the Levi-Civita part of the connection, 
responsible for the inhomogeneous transformations with respect to diffeomorphisms. Following the notations and conventions in \cite{Olmo:2022ops}, 
the contortion is defined as 
\begin{equation} 
    C_{~\ \sigma\mu}^{\rho} \equiv {\cal A}_{~\ \sigma\mu}^{\rho}-\Gamma_{~\ \sigma\mu}^{\rho}, 
\end{equation}
where $ {\cal A}_{~\ \sigma\mu}^{\rho}$ is the generic metric  connection, while $\Gamma_{~\ \sigma\mu}^{\rho}$ is its Levi-Civita component. Obviously, the torsion $ T_{~\ \sigma\mu}^{\rho}$ is identified with the antisymmetric part of the connection,
\begin{equation}  
    \label{torsion-distorsion}
     T_{~\ \sigma\mu}^{\rho} \equiv {\cal A}_{~\ \sigma\mu}^{\rho}-{\cal A}_{~\ \mu\sigma}^{\rho}
\end{equation}
and it is related to the contortion by
\begin{equation}  
    \label{contortion}
    2 \, C_{~\ \sigma\mu}^{\rho} \equiv  T_{~\ \sigma\mu}^{\rho} + T_{\sigma\mu}^{~\ ~\, \rho} + T_{\mu\sigma}^{~\ ~ \, \rho},
\end{equation}
so that a vanishing torsion implies an as well vanishing contortion. The curvature associated with ${\cal A}_{~\ \sigma\mu}^{\rho}$ is defined by 
\ \begin{equation} {\mathcal R}_{~\sigma\mu\nu}^{\rho} \equiv \partial_\mu{\cal A}_{~\sigma\nu}^{\rho}-\partial_\nu{\cal A}_{~\sigma\mu}^{\rho}+{\cal A}_{~\lambda\mu}^{\rho}{\cal A}_{~\sigma\nu}^{\lambda}-{\cal A}_{~\lambda\nu}^{\rho}{\cal A}_{~\sigma\mu}^{\lambda} 
\end{equation}
and can be expressed in terms of the contortion $C_{~\ \sigma\mu}^{\rho}$ as 
\begin{equation} 
    \label{FRC} 
     {\cal R}_{~\sigma\mu\nu}^{\rho}=R_{~\sigma\mu\nu}^{\rho} 
     +D_\mu C_{~\sigma\nu}^{\rho}-D_\nu C_{~\sigma\mu}^{\rho}+C_{~\lambda\mu}^{\rho}C_{~\sigma\nu}^{\lambda}-C_{~\lambda\nu}^{\rho}C_{~\sigma\mu}^{\lambda},   
\end{equation}
where $R_{~\sigma\mu\nu}^{\rho}$ is the ``standard''  Riemann tensor, depending solely on the Levi-Civita part of the connection. The curvature tensor can be contracted to provide the usual Ricci scalar curvature 
\begin{equation} 
    {\cal R} \equiv {\cal R}_{~~\mu\nu}^{\mu\nu}
\end{equation}
and a pseudoscalar 
\begin{equation} 
    {\cal R'} \equiv \varepsilon^{\mu\nu\rho\sigma}{\cal R}_{\mu\nu\rho\sigma},
\end{equation}
called the Holst invariant (see \cite{Hojman:1980kv,Nelson:1980ph,Holst:1995pc} for pioneering mathematical treatments
and \cite{Langvik:2020nrs,Karananas:2021gco,Gialamas:2022xtt,Gialamas:2024uar,Gialamas:2024iyu,Gialamas:2026pjo,Racioppi:2025igu,Katsoulas:2025srh} for specific inflationary universe application)
where $\varepsilon_{\mu\nu\rho\sigma}$ is the totally antisymmetric Levi-Civita tensor with 
$\sqrt{-g} \, \varepsilon_{0123}= 1$. It should be noticed that ${\cal R'}$ vanishes for $C_{~\sigma\mu}^{\rho}=0$, namely when the connection is the contortionless Levi-Civita one. This is the reason why, in the standard formulation of General Relativity (GR), ${\cal R'}$ is always absent.
However, it plays a prominent role in the class of theories that will be considered here, where the contortion is dynamical. 
In order to treat the coupling to fermions, it is convenient to resort to a first-order formulation of gravity. As customary (see appendix A for notations and conventions), the gravitational field is described by a vierbein $e^a_{~\mu}$ and a spin-connection $\omega^{a}_{~ \ b \mu}$ is introduced, that is a sum of a Levi-Civita component depending on the vierbein and the contortion.  In other words,
\begin{equation} 
    \omega^{a}_{~ \ b \mu} = \omega^{a}_{~ \ b \mu}(e) + C^{a}_{~ \ b \mu}.
\end{equation}
Of course, this is just a change of basis from the standard coordinate basis to the anholonomic (orthonormal) basis of the tangent bundle.  The relation between the coefficients of the connection in the two basis is simply given by the so-called first tetrad postulate. The relation between the curvature tensor in the two formulations is also quite simple, resulting in
\begin{equation} 
    {\cal R}_{~\sigma\mu\nu}^{\rho}=  e^{\rho}_{~ \, a} e^b_{~\, \sigma} {\cal R}_{~ b\mu\nu}^{a} \ ,
\end{equation}
where $e^{\mu}_{~ \, a}$ is the inverse vierbein. The scalar curvature ${\cal R}= e^{\mu}_{~ \, a} e^{\nu \,b} \ {\cal R}_{~\, b\mu\nu}^{a}$ and the Holst term ${\cal R'} =  e^{\mu}_{~ \, a} e^{\nu}_{~ \, b} \ \varepsilon^{a b}_{~ \, ~ \, c d} \, {\cal R}_{~\, ~ \, \mu\nu}^{c d}$ also have simple expressions, where $ \varepsilon_{a b c d}$, with  $ \varepsilon_{0123}=1$ is the Levi-Civita symbol in flat space.

In this paper, is of quite relevance the introduction of matter fields coupled to gravity, and a Palatini approach is convenient. As shown in \cite{Pradisi:2022nmh}, the models of interest can indeed be described using an action of the form 
\begin{equation} 
    S[e^{a}_{~ \mu},C,\Phi] = \int d^4x \, e \, \left[\alpha(\Phi)  {\cal R} + \beta(\Phi)  {\cal R}' + \Delta(\Phi, {\cal R}, {\cal R}')+ \Sigma(\Phi, {\cal D}\Phi, C)    \right], 
\label{curvLag}\end{equation}
where $e=\det{(e^{a}_{~ \mu})}$, $\Phi$ generically denotes all fields independent of the contortion entering the action through functions that respect  the (global and local) symmetries present in the Lagrangian. In particular, the $\alpha$ and $\beta$ functions are related to (possible) non-minimal couplings to the scalar and pseudoscalar curvatures, while $\Sigma(\Phi, {\cal D}\Phi, C)$ contains the ``matter'' fields and depends on the contortion both explicitly and through the covariant derivatives built out of the whole $\cal{A}$ connection. Finally, $\Delta$ is an arbitrary function of the indicated fields and curvatures carrying the non-linear terms. In \cite{DiMarco:2023ncs} it has been chosen to be
\begin{equation}
    \label{OurDelta}
    \Delta({\cal{R}'}) \, = \, \xi  \, {{\cal{R}}'}^p \ ,
\end{equation}
where $p>1$ is a real number and $\xi$ is a coupling constant with mass dimension $[m]^{4-2 p}$. It gives rise to an interesting set of inflationary models where the inflaton can be identified with a pseudoscalar field representing exactly a pseudo-scalar combination of the contortion components, thus originating directly and unequivocally from the geometry of the underlying Einstein-Cartan spacetime.  
To describe the inflationary scenario, it is convenient to take  preliminarily $\Sigma=0$, together with $2 \alpha(\Phi)= - M_{P}^2$ (thus directly the ``Einstein frame''), and  $4 \gamma  \beta(\Phi) = M_{P}^2$, where $\gamma$ is known as the Barbero-Immirzi parameter  \cite{BarberoG:1994eia,Immirzi:1996di}. 
As shown in \cite{Pradisi:2022nmh}, one may introduce an auxiliary pseudoscalar field $z$, in such a way that the previously defined class of  models is classically equivalent to
\begin{equation}
    S[e^{a}_{~ \mu},C,z]=\int d^4x \, e \, \left[- \frac{M_p^2}{2} {\cal R}+\left(\beta+\frac{\partial \Delta (z)}{\partial z }\right) {\cal R'} + \Delta(z) -z \frac{\partial \Delta (z)}{\partial z}\right] \ ,
\label{gravityAct}\end{equation}
provided $\frac{\partial^2\Delta}{\partial z^2}\neq 0$. Indeed, the equation of motion of the auxiliary field yields is $z={\cal{R}}'$, giving back (on shell) the previous model.  It is now an easy algebraic exercise to decompose the contortion into its irreducible components and to integrate them out.  Defining the quantity 
\begin{equation}
    B(z)= \frac{\beta+\frac{\partial \Delta(z)}{\partial z}}{M_P^2} \ ,
\end{equation}
it happens that its derivative sources the equations of motion of the vectorial and pseudovectorial components of the contortion. 
In other words, on shell the action can be written as 
the sum of the Einstein-Hilbert action and the lagrangian density of the pseudoscalar field $z$, 
\begin{equation}
\label{Eframeaction}
    S[e^{a}_{~ \mu},z] = \int d^4x\, e \, \left\{- \frac{M_P^2}{2} R+K(z)\frac{(D_{\mu} \, B(z))^2}{2}-V(z) \right\} , 
\end{equation}
where $R$ is the usual part of the scalar curvature that depends only upon the Levi-Civita spin connection, while 
\begin{equation} 
    \label{K-B relation}
    K(z)= \frac{24 M_p^2}{[1+16 B^2(z)]} 
\end{equation}
and the potential turns out to be
\begin{equation}
    V(z) =  z \frac{\partial \Delta(z)}{\partial z} - \Delta(z) .
\end{equation}
The action in Eq.\eqref{Eframeaction} suggests that $B(z)$ brings about the (non-canonical) kinetic term related to the pseudoscalaron $z$, which in turn is certainly not a ghost, since $K(z)$ is always positive.
Firstly, the field redefinition
\begin{equation}
    \phi(z) = \int^{z} d\zeta \sqrt{K(\zeta)} 
\end{equation}
allows to rewrite in a canonical way the action of the introduced pseudoscalar field $\phi$, being its kinetic term exactly the standard one.  The expression of $K(z)$ in terms of $B(z)$, allows us to establish the {\it universal} relation between the pseudoscalar field $\phi$ and $B(z)$, which holds for the whole considered class of models, {\it i.e.}

\begin{equation}
    \label{phivsz}
    \phi(z) - \phi_0 = \sqrt{\frac{3}{2}} \, M_P \, \sinh^{-1}{[4 B(z)]}, \quad 
\end{equation}
where $\phi_0$ is an integration constant.
Secondly, one needs to invert the previous relation to find $z$ as a function of $\phi$, in order to expose the potential $V=V[z(\phi)]$.
This procedure involves the solution of a complicated non-linear differential equation related to $\Delta(z)$ and its first derivative. 
In most cases, it is not possible to find an analytic solution. Fortunately, the simple form of the choice in Eq.\eqref{OurDelta} allows to write
the pseudoscalar sector of the action \eqref{Eframeaction} in terms of the canonically normalized field $\phi$.
Indeed, the potential can be written as
\begin{equation}
    V(z)= \xi (p-1) z^p 
\end{equation}
and Eq.\eqref{phivsz} can be explicitly inverted to give
\begin{equation}
    z^{p-1} = \frac{1}{\xi \, p} \left[ \frac{M_P^2}{4} \sinh{\left(\sqrt{\frac{2}{3}} \frac{1}{M_P} (\phi(z) - \phi_0) \right)} - \beta \right]  ,
\end{equation}
resulting in 
\begin{equation}
    V(\phi) = \frac{p-1}{p^{p/(p-1)}}\frac{1}{\xi^{\frac{1}{p-1}}}\Bigg|\frac{M^2_p}{4}
    \sinh\left( \sqrt{\frac{2}{3}}\frac{1}{M_p} (\phi - \phi_0) \right) - \beta \Bigg|^\frac{p}{p-1} .
\end{equation}
Thus, the effective cosmological action takes the standard form
\begin{equation}
    S[e^{a}_{~ \, \mu},\phi] \sim \int d^4x \, e \, \left( - \frac{M^2_p}{2} R + \frac{1}{2}\partial_{\mu}\phi\partial^{\mu}\phi - V(\phi)  \right),
\label{cosmoaction}
\end{equation}
where neither the background metric tensor $g_{\mu\nu}$ (or equivalently the vierbein) nor the integration constant $\phi_0$, 
which determines the pseudoscalaron vacuum expectation value, 
are constrained {\it a priori}.

\subsection{Inflationary phase and cosmological observations}
\label{Inflationary phase and cosmological observations}

\begin{figure*}[t]
    \centering
    \includegraphics[
      width=1\textwidth,
    ]{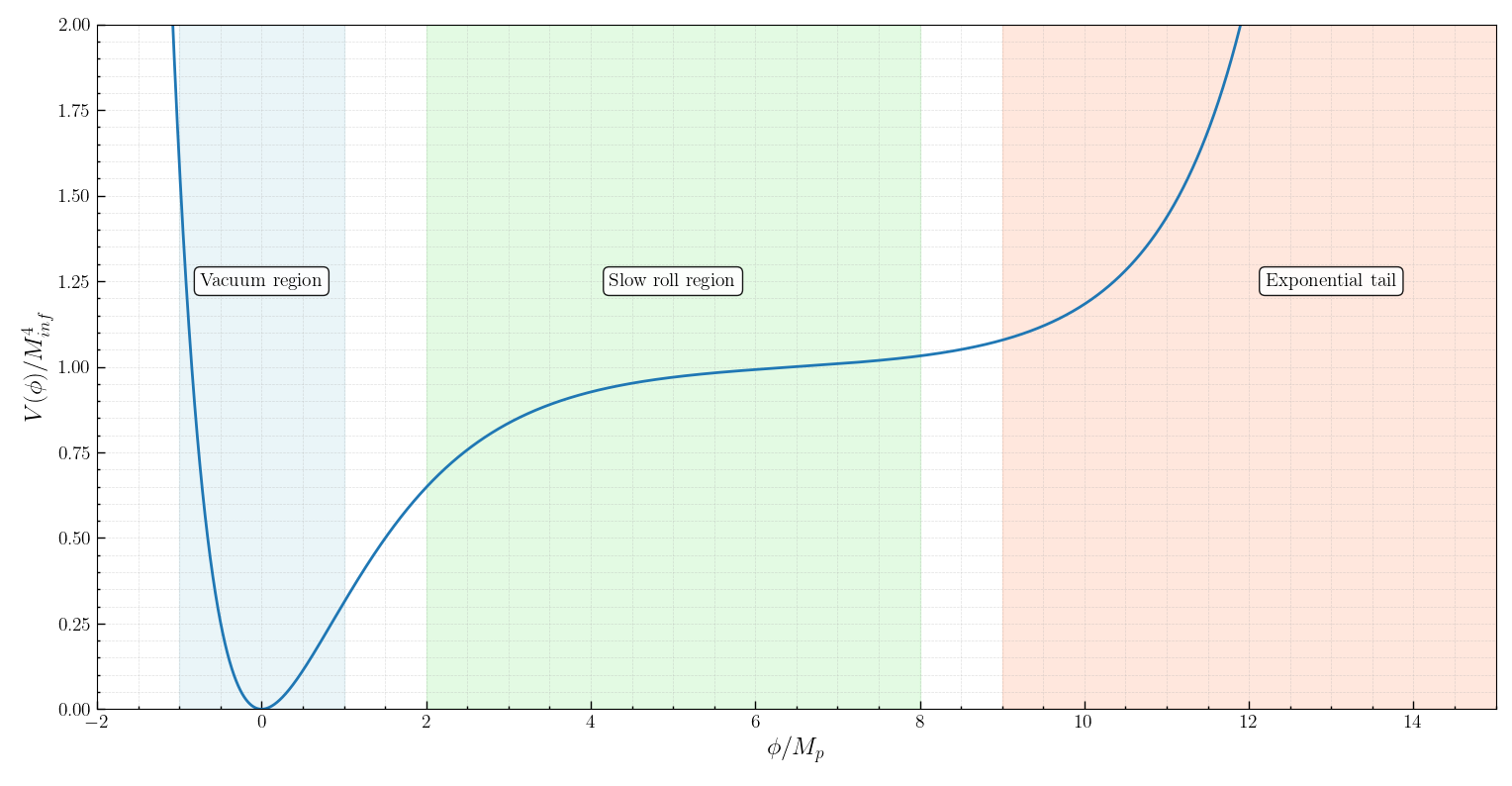}
    \caption{\it Inflationary potential $V(\phi)$, normalized to the reference inflationary scale $M_{\mathrm{inf}}^4$, 
    as a function of the rescaled scalar field $\phi/M_p$, shown here for the case $p=2$ and Barbero--Immirzi parameter $\gamma=-1/100$. In this scenario, the reference inflationary scale is assumed to be of order $M_{\text{inf}} \sim 10^{16}$ GeV. 
    The shaded vertical bands qualitatively highlight the three main regions (or regimes) of the model: 
    the vacuum region around the minimum of the potential, 
    the slow-roll region associated with the inflationary plateau, 
    and the large-field exponential tail.}
    \label{Fig:1}
\end{figure*}

The high energy cosmological action of Eq.\eqref{cosmoaction} can provide a standard period 
of slow-roll inflationary expansion by adopting two natural 
assumptions for the metric and the integration constant.
First, the background geometry can be approximated by a 
Friedmann–Lemaître–Robertson–Walker (FLRW) spacetime
\begin{equation}
    ds^2 \sim dt^2 - a^2(t)dl^2 ,
\end{equation}
where $dl$ is the line element of the three–dimensional (spatial) 
constant–time hypersurfaces, $t$ is the cosmic time and $a(t)$ is the 
dimensionless cosmic scale factor, allowing one to define the 
standard Hubble rate $H(t)=\dot{a}(t)/a(t)$.
Second, the integration constant $\phi_0$ can be chosen in such a way 
that the pseudoscalaron field $\phi$ is naturally interpreted as a 
scalar excitation oscillating around a flat Lorentz–invariant vacuum.
This requirement leads to
\begin{equation}
    \phi_0 = -\sqrt{\frac{3}{2}}M_p\sinh^{-1}{(\gamma^{-1})} ,
\end{equation}
that provides a vacuum expectation value located at $\phi= 0$, with $V(\phi=0) =  0$  
avoiding an additional Cosmological Constant term and allowing 
the scalar potential to take the convenient form
\begin{equation}
    V(\phi) = M^4_{\text{inf}} \, f_0(\phi) .
\label{eq:pseudopotential}\end{equation}
In Eq. \eqref{eq:pseudopotential} the inflationary reference scale is identified with
\begin{equation}
    M^4_{\text{inf}} = \frac{p-1}{p^{p/(p-1)}}\frac{1}{\xi^{\frac{1}{p-1}}}\Bigg|\frac{M^2_p}{4\gamma}\Bigg|^\frac{p}{p-1} ,
\end{equation}
while the dependence on the field is encoded in
\begin{equation}
    f_0(\phi) = \Bigg|\gamma\sinh X(\phi)-1\Bigg|^\frac{p}{p-1} ,
\end{equation}
where 
\begin{equation}
    X(\phi) = \sqrt{\frac{2}{3}} \frac{\phi}{M_p} + \sinh^{-1}(\gamma^{-1}) .
\end{equation}

The sign of the Barbero-Immirzi parameter determines the direction of the slow-roll phase.
Specifically, the slow-roll phase occurs for decreasing values of the inflaton field ({\it i.e.} $\dot{\phi}<0$)
for negative values of $\gamma$, 
while it occurs for increasing values of $\phi$
({\it i.e.} $\dot{\phi}>0$) for positive values of $\gamma$.
Moreover, the strength of the Barbero-Immirzi parameter controls height and shape of the inflationary potential, so that 
smaller values of $\gamma$ imply a higher inflation scale and a shorter plateau. 
The parameter $p$ also controls the extension of the inflationary plateau, the asymptotics of the potential for large field values and, most importantly,  
the vacuum geometry.
Indeed, as $p$ increases, the vacuum shape becomes more and more cuspy.

\begin{table}[t]
\centering
\small
\setlength{\tabcolsep}{4pt}
\renewcommand{\arraystretch}{1.1}

\begin{tabular}{p{0.53\textwidth} c c c}
\hline\hline
Dataset &
Scalar spectral index $n_s$ &
Tensor-to-scalar ratio $r$ &
$k_*^{(r)}$ \\
\hline

\multicolumn{4}{l}{\textbf{Planck + BICEP}}\\
\hline
\textit{Planck + lowE + lensing + BK15}~\cite{Planck:2018jri} 
& $0.9651 \pm 0.0041$ & <0.056 & 0.002\\
\textit{Planck + lowE + lensing + BK15 + BAO}~\cite{Planck:2018jri} 
& $0.9668 \pm 0.0037$ & < 0.058 & 0.002\\
\hline

\multicolumn{4}{l}{\textbf{Planck + ACT + DESI}}\\
\hline
\textit{Planck + ACT DR6 + lensing + DESI-DR1}~\cite{AtacamaCosmologyTelescope:2025blo} 
& $0.9743 \pm 0.0034$ & -- & --\\
\textit{Planck + ACT DR6 + lensing + DESI-DR2}~\cite{AtacamaCosmologyTelescope:2025blo} 
& $0.9752 \pm 0.0030$ & -- & --\\
\hline

\multicolumn{4}{l}{\textbf{Planck + ACT + SPT}}\\
\hline
\textit{Planck + ACT DR6 + SPT-3G-D1}~\cite{SPT-3G:2025bzu} 
& $0.9684 \pm 0.0030$ & -- & --\\
\textit{Planck + ACT DR6 + SPT-3G-D1 + DESI-DR2}~\cite{SPT-3G:2025bzu} 
& $0.9728 \pm 0.0027$ & -- & --\\
\hline

\multicolumn{4}{l}{\textbf{Planck + ACT + SPT + DESI + BICEP}}\\
\hline
\textit{Planck + ACT DR6 + SPT-3G-D1 + BK18}~\cite{Balkenhol:2025wms} 
& $0.9682 \pm 0.0032$ & < 0.034 & 0.05\\
\textit{Planck + ACT DR6 + SPT-3G-D1 + DESI-DR2 + BK18}~\cite{Balkenhol:2025wms} 
& $0.9728 \pm 0.0029$ & < 0.034 & 0.05\\

\hline\hline
\end{tabular}
\caption{Main constraints on the scalar spectral index $n_s$ (68\% CL) and the tensor-to-scalar ratio $r$ (95\% CL).
Planck refers to the latest Planck 2018 \textit{TT, TE, EE} measurements while $k_*^{(r)}$ labels the pivot scale in Mpc$^{-1}$,
for the tensor-to-scalar-ratio upper limit.}
\label{Tab:1}
\end{table}

\begin{figure*}[t]
    \centering
    \includegraphics[
      width=1\textwidth,
    ]{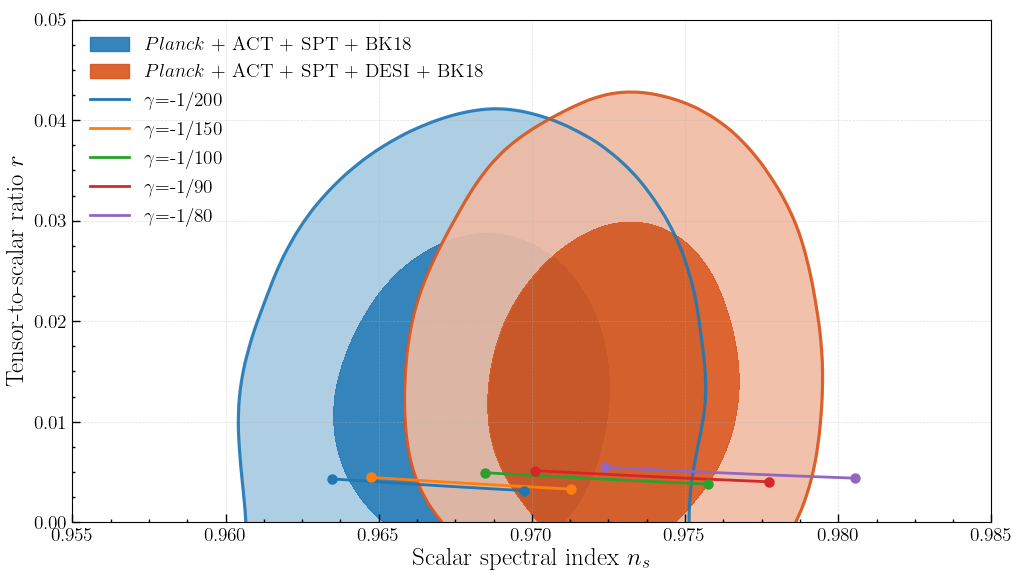}
    \caption{\it Predictions of the Einstein-Cartan-Holst  pseudoscalaron inflationary model ($p=2$) in the $(n_s, r)$ plane compared 
    with the $68\%$ and $95\%$ marginalized confidence regions derived from recent cosmological observations \cite{Balkenhol:2025wms}. 
    The blue region correspond to CMB-only constraints (Planck + ACT + SPT + BK18), 
    while the red region include the additional BAO information from DESI.}
    \label{Fig:2}
\end{figure*}

In this analysis, 
the focus is on the $p=2$ case (see Fig.\ref{Fig:1} for the corresponding potential shape).
In Tab. \ref{Tab:1} are reported the constraints on the main inflationary parameters, \textit{namely} 
the scalar spectral index $n_s$ and the tensor-to-scalar ratio $r$, provided by the latest CMB
(Planck, Atacama Cosmology Telescope (ACT), South Pole Telescope (SPT) and Bicep/Keck array experiments)
and baryon acoustic oscillation (BAO) missions (DESI), 
in several common combinations
\cite{Planck:2018jri,AtacamaCosmologyTelescope:2025blo,SPT-3G:2025bzu,Balkenhol:2025wms}.
In Fig. \ref{Fig:2} are shown the Einstein-Cartan pseudoscalaron inflationary predictions ($n_s,r$) 
for a set of Barbero-Immirzi parameter $\gamma$ values
and for a number of $e$-folds $N_e$ before the end of inflation \cite{DiMarco:2024yzn} such that
$50\leq N_e \leq 60$.

The predictions are compared with the corresponding $68\%$ and $95\%$ marginalized confidence regions 
from the CMB-based datasets, \textit{i.e.}
\textit{Planck + ACT DR6 + SPT-3G-D1 + BK18} \cite{Balkenhol:2025wms} and 
the full combination adding the BAO data from DESI, \textit{i.e.}
\textit{Planck + ACT DR6 + SPT-3G-D1 + DESI-DR2 + BK18} \cite{Balkenhol:2025wms}.
The model predictions display a systematic dependence on the Barbero-Immirzi parameter $\gamma$.
In particular, for $\gamma = -1/200$ and $\gamma = -1/150$, the predicted values of $(n_s,r)$ lie comfortably 
within the $68\%$ confidence region of the CMB-only dataset but still consistent
with the constraints obtained when BAO data from DESI are included. 
As $|\gamma|$ increases (\textit{e.g.} $\gamma = -1/100$ and $\gamma = -1/90$), 
the predictions shift toward larger values of $n_s$, moving closer to the central region favored by the combined CMB$+$DESI analysis
but are also extremely compatible with the tighter constraints from CMB-only dataset.
The case $\gamma = -1/80$ is consistent at the $95\%$ confidence level with the CMB dataset
and within the tighter marginalized region obtained with BAO. 
In all cases, the predicted tensor-to-scalar ratio remains safely below the current upper limits.
As shown in \cite{DiMarco:2023ncs} the $|\gamma|\sim 10^{-2}$ provides a high-energy inflation scale
with reference parameter $M_{\text{inf}}\sim 10^{16}$ GeV and self-coupling of pseudoscalaron curvature\footnote{Note that, despite the large $\xi$ value, the term in Eq. \eqref{OurDelta} just  represents a small non-linear correction to the Einstein-Hilbert term in the gravity action, and correctly fixes the scale of inflation in the dual pseudoscalar-tensor description.} $\xi\sim 10^9$.

It is important to conclude this section with a remark concerning the datasets employed in the analysis.
The measurements from the CMB experiments, \textit{i.e.} Planck, ACT, SPT,
are mutually consistent and show no statistically significant evidence 
for deviations from the standard $\Lambda$CDM framework. 
In particular, the constraints on the scalar spectral index $n_s$ derived from these CMB experiments are fully compatible between each others.
Similarly, the latest DESI DR2 BAO measurements alone are consistent with $\Lambda$CDM. 

However, there is a mild but non-negligible statistical discrepancy between CMB and DESI DR2 BAO constraints \emph{within} $\Lambda$CDM.
This so-called \emph{BAO-CMB tension} \cite{Ferreira:2025lrd,Ye:2025ark} 
corresponds to a $2\sigma$ - $3\sigma$ mismatch between BAO and CMB constraints for the couple $(\Omega_m, r_d h)$,
where $\Omega_m$ is the present-day matter density fraction, $r_d h$ is the sound horizon at the baryon drag epoch,
and $h = H_0 / (100\,\mathrm{km\,s^{-1}\,Mpc^{-1}})$ is the dimensionless Hubble parameter.
Moreover, the statistical significance of the tension also appears to be sensitive to the inclusion of different Type Ia supernova datasets.
This tension could potentially hint at several extensions of $\Lambda$CDM, including dynamical dark energy \cite{Ye:2025ark},
or could arise from some systematic effects or features of the data analysis pipeline.
In this context, one could wonder about the \emph{robustness} of using a combination of CMB and BAO datasets 
to provide \emph{reliable} confidence level for cosmological and inflationary parameters.
The current level of tension is not sufficient to claim for a dramatic breakdown of $\Lambda$CDM
(eventually challenged by other observations)
but it is important to stress this point when discussing joint constraints.

\subsection{The physics around the vacuum}
\label{The physics around the vacuum}

The cosmological inflationary phase ends when the inflaton field reaches 
the model slow-roll breaking value $\phi_{\rm end}$, after which it undergoes a relaxation toward the minimum of the scalar potential at $\phi = 0$.
The geometry of the vacuum can be explored by series expanding the scalar potential, {\textit{e.g.}} up to the fourth order, as
\begin{eqnarray}
    V(\phi)\simeq \frac{m^2_{\phi}}{2}\phi^2
    + \frac{g_{\phi}}{3!}\phi^3 + \frac{\lambda_{\phi}}{4!}\phi^4, 
\end{eqnarray}
where the coefficients $m_\phi$, $g_\phi$, and $\lambda_\phi$, given by
\begin{equation}
    m_\phi^{2}\equiv V''(\phi)\Big|_{\phi=0},\quad
    g_\phi\equiv V'''(\phi)\Big|_{\phi=0},\quad
    \lambda_\phi\equiv V^{(IV)}(\phi)\Big|_{\phi=0},
\end{equation}
represent the mass of the inflaton excitation and the effective (cubic and quartic) self-coupling of the pseudoscalar field around the vacuum, respectively. In principle, they depend on the Barbero-Immirzi parameter. However, in a CMB-prediction consistent regime, where $|\gamma|\ll 1$, 
one easily gets
\begin{align}
m_\phi^{2}
&\simeq2M_{\mathrm{inf}}^{4}\left(\frac{b}{M_p}\right)^{2} + \mathcal{O}(\gamma^2),\\
g_\phi
&\simeq6M_{\mathrm{inf}}^{4}\left(\frac{b}{M_p}\right)^{3}\mathrm{sgn}(\gamma) + \mathcal{O}(\gamma^2),\\
\lambda_\phi
&\simeq14M_{\mathrm{inf}}^{4}\left(\frac{b}{M_p}\right)^{4} 
+ \mathcal{O}(\gamma^2),
\end{align}
where $b^2=2/3$. Supposing an inflationary reference scale $M_{\text{inf}}\sim10^{16}$ GeV 
(see Sec. \ref{Inflationary phase and cosmological observations}), 
one finds 
\begin{equation}
    m^2_{\phi}\sim 10^{28} \mbox{ GeV}, \quad g_{\phi}\sim -2\times 10^{10} \mbox{ GeV}, \quad \lambda_{\phi}\sim 3\times 10^{-8}.
\end{equation}

Therefore, the inflaton scalar mass would be of order $10^{14}$ GeV - 
a relatively high value if compared to the ones of other inflationary models -
and constitutes the most relevant term of the vacuum geometry.
The evolution of the inflaton field around the vacuum is governed by the standard equation of motion
\begin{equation}
\ddot{\phi} + 3H\dot{\phi} + \Gamma_\phi\,\dot{\phi} + V'(\phi) = 0 \, ,
\end{equation}
which describes the oscillatory dynamics of the homogeneous inflaton condensate.
As it is well known, the Hubble friction term ($3H\dot{\phi}$) is related to the expansion of the Universe and 
leads to a decrease of the oscillation amplitudes,
with a time dependence determined by the background energy density and by the effective equation of state of the postinflationary cosmic fluid.  The phenomenological term ($\Gamma_\phi\,\dot{\phi}$), on the other hand,  accounts for the decay of the inflaton into lighter SM or BSM particles, 
further introducing an additional source of dissipation,
whose magnitude depends on the microscopic details of the underlying gravity and particle physics models. 
Finally, the ``force'' term ($V'(\phi)$) is dominated 
by the linear contribution coming from the effective (quadratic) mass term, while the higher-order self-interactions 
give only rise to subleading anharmonic corrections.
The postinflationary dynamics with the associated processes can be obtained, in principle, 
by solving the system of coupled equation of motions of the inflaton  
and the remaining matter fields. 
However, a description based on evolving perfect fluids provides 
a more convenient framework for capturing the macroscopic features of the reheating phase.

The corresponding set of integrated Einstein-Boltzmann equations for the energy and number densities will thus be adopted in the next section.

\section{Reheating and nonthermal leptogenesis}
\label{Reheating and nonthermal leptogenesis}

In this section, the structure of the postinflationary gravity - BSM matter lagrangian and its cosmological consequences is analyzed in detail.
The coupling of the gravitational (pseudoscalaron) sector to additional fundamental fields and, in particular, 
to generic fermion fields is first examined, emphasizing its universal nature. 
Then, the analysis specifically focuses on the coupling of the pseudoscalaron to a type-I seesaw extension of the Standard Model with three sterile hierarchical Majorana RHNs which, in turn, interact with the SM fermions through the Higgs sector.

\subsection{Einstein-Cartan pseudoscalaron coupling to fermions}
\label{Pseudoscalaron interactions with fermions}

In the Einstein-Cartan-Holst framework, the interaction between the gravitational sector and matter fields
is obtained by specifying a suitable SM or BSM matter term $\Sigma$ in the action of Eq. \eqref{curvLag}. 
As mentioned, the matter action contains covariant derivatives 
that implement minimal couplings to several SM or BSM fundamental fields. 
However, it is well known that minimally coupled scalar and vector fields do not feel the contortion \cite{Hehl:1976kj}.  
As a consequence, no minimal interactions between inflaton and scalar or gauge fields will arise.
By contrast, the minimal coupling of the gravitational sector to a Dirac
(or Majorana\footnote{Majorana fermions satisfy $\psi^c=\psi$. For Weyl fermions, a bare mass term
cannot be present. In chiral theories, like the SM, Dirac masses arise from the Higgs mechanism.})
fermion field $\psi$ of mass $m$ can be obtained by considering a matter term of the form (see Appendix A)
\begin{equation}
    \mathcal{S}_{f} =  \int d^4x \,  e \,  S(\psi, { \mathcal{D}}\psi, C) = \int d^4x \, e \, \zeta \, \left[ \frac{i}{2} \, \left(\overline{\psi} \, \gamma^{\mu}\,   \mathcal{D}_{\mu} \, \psi - \overline{ \mathcal{D}_{\mu} \psi} \ \gamma^{\mu} \, \psi \right) - m \overline{\psi} \psi \right] , 
\end{equation}
where $\zeta$ is $1$ ($1/2$) for Dirac (Majorana) spinors.
The gamma matrices in curved spacetime are defined as $\gamma^{\mu}=e^{\mu}_{~ \, a} \, \gamma^{a}$
(with flat Latin indices), while the covariant derivative contains the spin connection. An important observation is that the contortion in the non-holonomic basis is also related in the obvious way,  $C^{a}_{~ \ b \mu}\, = \, e^{a}_{~ \,\rho} \, e^{\sigma}_{~ \, b} \, C^{\rho}_{~ \ \sigma \mu} $, to the one in Eq. \eqref{contortion}. 
Using it in the expansion of covariant derivatives makes the fermion--contortion interaction explicit. One gets 
\begin{equation}
    \mathcal{L}_{int} = \frac{\zeta}{4} \, \varepsilon^{a b c d} \, C_{a b c} \, \bar\psi \, \gamma_{d} \, \gamma^5 \,  \psi,
\end{equation}
recovering the classical result stating that only the totally antisymmetric part of the contortion 
couples to fermions in Einstein-Cartan gravity \cite{Hehl:1976kj,Shapiro:2001rz,Hammond:2002rm,Karananas:2021zkl}. As expected, it results in a linear coupling between the pseudovectorial component $a^d$ of the torsion\footnote{The pseudovectorial component of the torsion is defined to be $a_{\mu} = \varepsilon_{\mu \nu \rho \sigma} \, T^{\nu \rho \sigma}$, see \textit{e.g.} \cite{Shapiro:2001rz}.}, dual to the pseudoscalaron field,
and the pseudovector bilinear Dirac (or Majorana) term $\ell_{d} \equiv \bar\psi \, \gamma_{d} \, \gamma^5 \,  \psi$. 
Indeed, this is the reason why in theories with gravity linear in the curvatures the contortion satisfies algebraic equations, being thus non-dynamical. 
In the present case, the coupling enters the connection equations of motion as an additional source term that combines with the kinetic term of the pseudoscalaron.
Going on-shell by integrating out the contortion (like in the absence of fermions) one gets
\begin{equation}
  \mathcal{S}_{f} = \int d^4x \, e \,  \left[ \frac{i \, \zeta }{2} \, \left(\overline{\psi} \, \gamma^{\mu}\,   D_{\mu} \, \psi - \overline{D_{\mu} \psi} \ \gamma^{\mu} \, \psi \right) - m \, \zeta \, \overline{\psi} \psi   + \frac{\mathcal{C}_{\phi \psi \psi}}{M_P} \, \partial_{\mu}\phi \, \left( \bar\psi  \, \gamma^{\mu}\, \gamma^5 \, \psi \right) \, + \, \frac{\mathcal{C}_{4 \psi}}{M_P^2} \, \left( \bar\psi  \,  \gamma^{\mu}\, \gamma^5 \,  \psi \right) \, \left( \bar\psi  \, \gamma_{\mu}\, \gamma^5\,  \psi \right) \right] .
\label{intferm}
\end{equation}
The first two terms are the standard kinetic and mass terms of a Dirac (or Majorana) fermion, involving just the Levi-Civita covariant derivative.
The third term represents the \emph{universal} coupling of the pseudoscalaron to a generic Dirac (or Majorana) fermion, 
\begin{equation}
     \mathcal{C}_{\phi \psi\psi} = \frac{3\, \zeta \, }{1 +16 B^2} \, \frac{\partial B(\phi)}{\partial \phi}, 
\end{equation}
while the last one represents an additional effective term quartic in the fermions, 
\begin{equation}
     \mathcal{C}_{4\psi} = \frac{3\, \zeta^2 \, }{16 \,(1 +16 B^2)} , 
\end{equation}
very familiar both from Einstein-Cartan gravity and from supergravity \cite{Freedman:2012zz}.
The quartic term is naturally suppressed with respect to the cubic coupling, already tiny because of the presence of the inverse reduced Planck mass. In order to obtain the decay rate of the inflaton, the previous coefficients must be evaluated around the minimum of the potential, \textit{i.e.} for $B$ and its derivatives at $\phi=0$, 
where they can be written as
\begin{equation}
     \mathcal{C}_{\phi \psi\psi}  = \sqrt{\frac{3}{8}} \, \frac{\zeta \, \gamma}{\sqrt{1+\gamma^2}}  
\label{3plecoupl}
\end{equation} 
and
\begin{equation}
    \mathcal{C}_{4\psi} = - \frac{3}{16} \, \frac{\zeta^2 \, \gamma^2}{1 + \gamma^2}  .
\end{equation}
Not surprisingly, they are both singular for $\gamma=\pm i$, 
values at which the Holst and the Einstein-Hilbert scalar terms combine to give the contracted (anti)self-dual curvature two-form.  
The coupling of Eq.\eqref{3plecoupl} can be used to compute the  \emph{universal} decay rate at zero temperature of the inflaton to a generic pair of fermions $\psi$. To this, it is useful to recall that the contribution of the Lorentz invariant phase space
for a $(1 \to 2)$ is given by \cite{Peskin:1995ev}
\begin{equation}
 R_2(s) = \frac{1}{(2 \pi)^2} \int \frac{d^3p_1}{2E_1} \frac{d^3p_2}{2E_2} \delta^{(4)}(k - p_1 - p_2) .
\end{equation}
In the frame where the decaying particle is at rest, $k^{\mu} = (\sqrt{s}, \vec{0}) = (m_{\phi},  \vec{0}) $, 
one gets a factor 
\begin{equation}
    R_2(s)= \frac{\zeta}{8 \pi} \sqrt{1 - \frac{4 m_{\psi}^2}{m_{\phi}^2}} 
\end{equation}
for Dirac (Majorana) fermions, to be multiplied by the contribution coming from the matrix of the process in the momentum space,  $\mathcal{M}$. 
Here, apart from the coefficient in Eq. \eqref{3plecoupl} and taking into account the derivative coupling, the matrix element is
\begin{equation}
    \mathcal{M} = \frac{k^{\mu}}{M_P} \bar{u}(p_1) \gamma_{\mu}\, \gamma^5 v(p_2) ,
\end{equation}
where $u$ and $v$ are the usual positive- and negative-frequency spinors in the momentum space, normalized as in \cite{Peskin:1995ev}. 
To get the unpolarized decays, one has to sum the modulus squared of the amplitude components over the final states. Using standard techniques related to the trace of products of gamma matrices, one gets a contribution 
\begin{equation}
    X = \frac{1}{2} \sum_{r,s} | \mathcal{M}_{r,s}|^2 = 4 m^2_{\phi} m^2_{\psi} . 
\end{equation}
Thus, including the normalization factor $(2 m^2_{\phi})^{-1}$ of the initial particle, the final expression of the decay rate of the pseudoscalaron to a Dirac (Majorana) fermion $\psi$ turns out to be
\begin{equation}
    \Gamma_{\phi\rightarrow\psi\psi} = |C_{\phi\psi\psi}|^2 \frac{\zeta^3 \, m_{\phi}m^2_{\psi}}{4 \pi M^2_p}\sqrt{1 - \left(\frac{2 m_{\psi}}{m_{\phi}}\right)^2} =\frac{3 \, \zeta^3\,m_{\phi} \, m^2_{\psi}}{32\, \pi\, M^2_p}\, \frac{\gamma^2}{1 + \gamma^2} \, \sqrt{1 - \frac{4 m_{\psi}^2}{m_{\phi}^2}} \ .
\label{UnivDecRate}
\end{equation}

Therefore, the inflaton decay rate in Eq.~\eqref{UnivDecRate} scales with the square of the fermion mass. 
As a consequence, inflaton decays into some heavy BSM fermions,
such as heavy right-handed neutrinos
are, in a natural way, largely favorite with respect to decays into lighter fermions, such as those of the Standard Model.
This remains true both in the unbroken electroweak phase - where SM fermions are effectively massless - 
and in the broken phase, where the heaviest SM state is the top quark ($\sim 171$ GeV).

As a result, a reheating driven by the coupling of the inflaton field to a (SM-coupled) heavy RHN is inevitably characterized by a nonthermal leptogenesis mechanism.
In the following section, an explicit pseudoscalaron–type-I-seesaw model is constructed 
and then used to derive the resulting lepton asymmetry generation.

\subsection{Einstein-Cartan pseudoscalaron coupling to seesaw type I model}
\label{Coupling to the seesaw Type I model}

The cosmological observations reveal a fundamental matter-antimatter asymmetry in the Universe
known as baryon asymmetry and currently constrained to be\footnote{Strictly speaking, the baryon asymmetry is defined as $\eta_B \equiv (n_B - n_{\bar B})/s$. 
After baryon-antibaryon annihilation one has $n_{\bar B} \ll n_B$, so that $n_B - n_{\bar B} \simeq n_B$, 
and the asymmetry is commonly expressed as $n_B/s$.} $n_B/s \sim 8.7\times10^{-11}$
\cite{Fields:2019pfx,Cooke:2017cwo,ParticleDataGroup:2024cfk}.
Such an imbalance can be generated dynamically by $B$-, $C$-, and $CP$-violating 
interactions taking place during some out of equilibrium phase in the early Universe \cite{Sakharov:1967dj}.
The SM contains all these processes
-- $B$ violation from anomalies, 
and $C$ and $CP$ violation from weak interactions through chirality and the complex phase of the CKM matrix -- 
but the predicted asymmetry is far below the observed one, thereby indicating the need for physics beyond the SM.
In this context, one of the most appealing ways to produce a matter-antimatter asymmetry
is through leptogenesis, where an early asymmetry in the leptonic sector is converted into 
baryon asymmetry via the nonperturbative electroweak sphaleron transitions.
The leptogenesis mechanism can be naturally embedded in several extensions of the SM, some 
designed to also address other cosmological and particle physics puzzles. 
In particular, type I seesaw leptogenesis provides an appealing  framework for both the generation
of small SM-doublet neutrino masses and lepton asymmetry. 
Indeed, the addition of a certain number $k\ge2$ of sterile RHNs with large bare Majorana masses $M_{N_k}$ naturally provides tiny masses for the resulting active left-handed neutrinos (the seesaw mechanism). 
Moreover, assuming a mass hierarchy among RHNs and neglecting flavour effects makes the production of lepton asymmetry 
typically driven by the lightest RHN state, here indicated as $N_1$.
Leptogenesis can be thermal or nonthermal -- depending on the way the heavy Majorana RHNs are produced --
and the latter is especially attractive, as it can help to evade cosmological bounds on the mass of the lightest state 
as well as to alleviate issues related to overproduction of gravitinos in supergravity scenarios.
Most analyses of nonthermal leptogenesis have been developed for a purely scalar inflaton with 
standard Yukawa couplings to RHNs, in a General Relativity background.
In the present case, on the contrary, the focus will be on a
postinflationary Einstein-Cartan pseudoscalaron–-type-I--seesaw setup, investigating its ability to guarantee
a robust phase of nonthermal leptogenesis 
with consistent predictions for baryon asymmetry 
through the distinctive interactions encoded in Eq. \eqref{intferm}.
In light of this, 
the starting point is adding to Eq. \eqref{cosmoaction} a postinflationary (non-supersymmetric) type I seesaw matter lagrangian (with three RHNs $N_k$) of the form

\begin{equation}
\Sigma = 
\mathcal{\widetilde{L}}_{SM}
+ \sum_{k=1}^3 S(N_k, \mathcal{D}N_k, C)
- \left[
\sum_{i=1}^3 \sum_{k=1}^3  {\cal Y}_{ik}\, (\overline{\psi}_{\ell_i} \cdot \tilde h)\, N_k
+ \frac{1}{2} \sum_{j=1}^3 \sum_{k=1}^3 (M_N)_{jk}\, \overline{N^c_j}\, N_k
+ \text{(h.c.)}
\right] \, .
\label{mattlagrinter}
\end{equation}
Here, $\mathcal{\widetilde{L}}_{SM}$ is the Standard Model lagrangian containing covariant derivatives (with spin connection) of the various fields, $S(N_k,{ \mathcal{D}}N_k, C)$ is the kinetic lagrangian of the $N_k$ RHN,  adapted from Eq. \eqref{intferm} and containing the corresponding couplings to the pseudoscalaron, 
while the third term encodes the type I seesaw sector. It contains the sum of two distinct contributions:
the first one consists of the interactions between the SM leptons ${\psi}_{\ell_i}$, the Higgs field $h$\footnote{$\tilde{h}= i \sigma_2 h^*$ is the conjugate Higgs field.} and the RHNs. The complex Yukawa couplings ${\cal{Y}}_{i k}$ give rise to the Dirac mass matrix after electroweak symmetry breaking, with
\begin{equation}
   (m_D)_{i k} = \frac{v \, {\cal{Y}}_{i k}}{\sqrt{2}} ,
\label{diracmassmat}\end{equation}
where $v\sim 246$ GeV is the Higgs field vacuum expectation value. 
The second contribution
is the bare Majorana mass matrix of RHNs, 
taken diagonal in the basis of RHN mass eigenstates $M_N= diag(M_{N_1},M_{N_2}, M_{N_3})$\footnote{The presence of a Majorana (bare or effective) mass matrix for the left-handed SM neutrinos is  in principle admissible but excluded in this paper.}. 
Using a basis of left-handed fermions, $\underline{\nu} = (\nu_L^i, {N_R^c}^{k})^T$, the effective lagrangian mass term of neutrinos can be written in the convenient (Majorana) form
\begin{equation}
    \mathcal{L}_{mass} = - \frac{1}{2} \ \overline{\underline{\nu}^c} \ \mathcal{M}_{\nu} \ \underline{\nu} \ + \ h.c.
\end{equation}
where  the complete mass matrix 
\begin{equation}
    \mathcal{M}_{\nu} = \begin{pmatrix}
0 & m_D \\
m_D^T & M_N
\end{pmatrix}
\end{equation}
turns out to be complex and symmetric, due to the flip properties of fermion bilinears. It can be diagonalized using a unitary matrix $V$.
Assuming large Majorana masses -- \textit{i.e.} such that the diagonal elements of $M_N$ are much larger than the ``electroweak scale'' characterizing the elements of $m_D$ in Eq. \eqref{diracmassmat} -- 
the diagonal states of the complete mass matrix are three light (almost active) 
Majorana neutrinos of masses of order $m_{\nu} \sim (m_D M_N^{-1} m_D^T)$, and three heavy (almost sterile) Majorana neutrinos of masses of order $M_N$. As the charged leptons in Eq. \eqref{mattlagrinter} are chosen in the mass eigenstate basis, the leptonic mixing matrix entering the charge current interactions coincides with the matrix $U$ that defines the light neutrino mass eigenstates. In the considered case, it is a $3 \times 3$ matrix analogous to the 
CKM matrix of the quark sector, usually indicated as PMNS matrix\footnote{To be precise, the so called PMNS matrix \cite{Pontecorvo:1957qd} is the one that enters the extension of the SM where the three sterile RHNs are simply the right-handed massless partners 
of left-handed neutrinos of the SM, and can be parameterized just using three angles and a (Dirac) phase. By analogy, the same name is used in the more general case of $k$ added (massive) sterile RHNs.} 
and can be parameterized in terms of three angles, one Dirac phase, and two Majorana phases. 
The complete model contains many unknown parameters (21 in the type I seesaw with three RHNs, apart from the SM ones) 
which cannot be completely fixed by fitting experimental data. 
Considering simpler situations, for instance adding just two RHNs \cite{Minkowski:1977sc} 
(thus keeping one of the active neutrinos massless), would reduce the number of unknown parameters.
Those cases will not be pursued in this paper where, in any case, the interest resides in showing that a wide range of compatibility with the experimental data is available for the decisive Barbero-Immirzi parameter.

In order to get active neutrino masses compatible with experimental limits, $\sum m_{\nu}\lesssim 7\times 10^{-2}$ eV (see for instance
\cite{DESI:2024mwx,Allali:2024aiv})
, the ratio between the modulus squared of the Yukawa couplings and the RHN mass must be properly tuned. Further simplifications are obtained by assuming a hierarchical RHN spectrum, for instance $M_{N_1} \ll M_{N_2},\, M_{N_3}$, with the 
inflaton decay to the lightest state \emph{as the only one} kinematically allowed,  
\textit{namely} $M_{N_1} \lesssim m_{\phi}/2 \ll M_{N_2},\, M_{N_3}$. Under these conditions, the postinflationary dynamics is governed by an open decay channel of the inflaton into
$N_1$ pairs, while the heavier states cannot be produced. 
As a result, $N_2$ and $N_3$ do not play any \emph{explicit} cosmological role in the reheating dynamics.
The corresponding decay width of the inflaton into the lightest right handed neutrinos, adapted from Eq.\eqref{UnivDecRate}, is
given by
\begin{equation}
\Gamma_{\phi \to N_1 N_1} = \frac{3\,m_{\phi}\,M_{N_1}^2}{256\,\pi\,M_p^2}\,
\frac{\gamma^2}{1+\gamma^2}\,
\sqrt{1-\frac{4M_{N_1}^2}{m_{\phi}^2}} \, ,
\end{equation}
while the decay channel into massless leptons gets practically suppressed.
When produced, $N_1$s then decay into Standard Model radiation (leptons and Higgs bosons), with a decay width given by
\begin{equation}
\Gamma_{N_1 \to RR} = \frac{[{\cal Y}^\dagger {\cal Y}]_{11}\,M_{N_1}}{8\pi}\, .
\label{NdectoRR}
\end{equation}
It is crucial to understand the order of magnitude of $N_1$ decay to radiation. To this aim, its decay width is
often expressed in terms of $\tilde m_1$, the so called  \emph{effective}, or
\emph{reference}, light neutrino mass,  in the form
\begin{equation}
\Gamma_{N_1\to RR}
=
\frac{\tilde m_1\, M_{N_1}^2}{8\pi v^2}\, .
\end{equation}
It correctly describes the \emph{true} decay width and
lifetime of the RHN. The $\tilde m_1$, measuring the size of the ratio of Dirac to Majorana mass, is a very relevant quantity,  
most commonly employed in the Boltzmann treatment of leptogenesis \cite{Plumacher:1996kc,Buchmuller:2004nz}.
The decay of $N_1$ eventually drives the generation of lepton asymmetry. 
In particular, the RHN mass, together with the heaviest light neutrino mass, also bound the
magnitude of the $CP$ asymmetry parameter $\varepsilon_1$ \cite{Davidson:2002qv}
(the parameter that controls the lepton asymmetry production, see Appendix \ref{appB})
\begin{equation}
|\varepsilon_1| \lesssim \frac{3}{16\pi}\,
\frac{M_{N_1}\, m_\nu^{\rm max}}{v^2}\, .
\end{equation}
A few numbers help to illustrate the situation.
Given an RHN sector with $M_{N_1}\sim 10^{13}$ GeV and an
effective light-neutrino mass (which can also be close to the constrained $m_{\nu}^{\text{max}}$)
$\tilde m_1\sim 10^{-3}$ eV, the corresponding RHN decay width is $\Gamma_{N_1 \to RR}\sim 7\times 10^{7}\ {\rm GeV}$. 
In the CMB-favoured scenarios with $p=2$ and
$\gamma\sim -1/100$, the inflaton mass is
$m_\phi\sim 10^{14}$ GeV (see Sec.~\ref{The physics around the vacuum}),
which implies an inflaton decay width of order
$\Gamma_{\phi \to N_1N_1}\sim 7\times 10^{-3}$ GeV.
Therefore, $\Gamma_{\phi \to N_1N_1}/\Gamma_{N_1 \to RR}\sim 10^{-10}$,
showing that the RHN decay is essentially instantaneous on the
timescale set by inflaton decay.
In this respect, nonthermal leptogenesis is primarily driven by
the inflaton decay, which continuously produces RHN particles
that then decay rapidly into a relativistic SM plasma carrying
a net lepton asymmetry.
It is also worth stressing that, if the heaviest SM neutrino state alone determines the 
mass sum $\sum m_\nu$, \textit{i.e.}
$m_\nu^{\rm max}\sim 7\times 10^{-2}$ eV, the Davidson--Ibarra bound
yields an upper limit on the $CP$ asymmetry parameter of order
$|\varepsilon_1|\lesssim 7\times 10^{-3}$. This value provides a
useful benchmark for the dynamics: scenarios requiring larger
values of $\varepsilon_1$ for reproducing the observed
baryon asymmetry should be regarded as disfavoured.

\subsection{Boltzmann equations : numerical and analytical solution}

In order to study in details the post-inflationary evolution of the Einstein-Cartan pseudoscalaron scenario,
it is crucial to solve the Einstein-Boltzmann equations describing reheating and nonthermal leptogenesis mediated by the decay of 
the lightest RHN $N_1$ of the seesaw sector.
The simplest version of the Einstein-Boltzmann system can be written as
\begin{eqnarray}
\dot{\rho}_\phi(t) + 3 H(t)\,\rho_\phi(t)
&=& - \Gamma_{\phi \to RR}\,\rho_{\phi}(t) - \Gamma_{\phi \to {N_1}{N_1}}\,\rho_{\phi}(t) , \\
\dot{\rho}_{N_1}(t) + 3 H(t)\,\rho_{N_1}(t)
&=& \Gamma_{\phi\to {N_1}{N_1}}\,\rho_\phi(t) - \Gamma_{{N_1}\to RR}\,\rho_{N_1}(t) ,\\
\dot{\rho}_R(t) + 4 H(t)\,\rho_R(t)
&=& \Gamma_{\phi\to RR}\,\rho_{\phi}(t) +\Gamma_{{N_1}\to RR}\,\rho_{N_1}(t) , \\
\dot{n}_{L}(t) + 3 H(t)\,n_{L}(t) &=&  \varepsilon_1 \ \Gamma_{{N_1}\to RR}\ \frac{\rho_{N_1}}{M_{N_1}}
\end{eqnarray}
where $\rho_\phi$ denotes the inflaton energy density, $\rho_{N_1}$ the energy density of the $N_1$ right-handed neutrino, $\rho_R$ the radiation energy density and $n_L$ the lepton asymmetry number density, $n_L = n_{\ell} - n_{\bar\ell}$. 
The evolution is parameterized by cosmic time $t$ and the Hubble rate is determined by the first Friedmann equation
\begin{eqnarray}
H^{2}(t) = \frac{1}{3M^2_p}\left[
\rho_{\phi}(t) + \rho_{N_1}(t) + \rho_R(t)
\right].
\end{eqnarray}
The initial conditions, given by
\begin{eqnarray}
\rho_{\phi}(t_{\text{end}}) = \rho(\phi_{\text{end}}), \quad
\rho_{N_1}(t_{\text{end}}) \simeq 0, \quad
\rho_R(t_{\text{end}}) \simeq 0, \quad
n_{L}(t_{\text{end}}) \simeq 0 ,
\end{eqnarray}
correspond to an inflaton-dominated configuration at the end of inflation, while all other components, as well as any pre-inflationary lepton asymmetry, are strongly suppressed by the preceding accelerated expansion.

\begin{figure*}[t]
    \centering
    \includegraphics[
      width=1\textwidth,
    ]{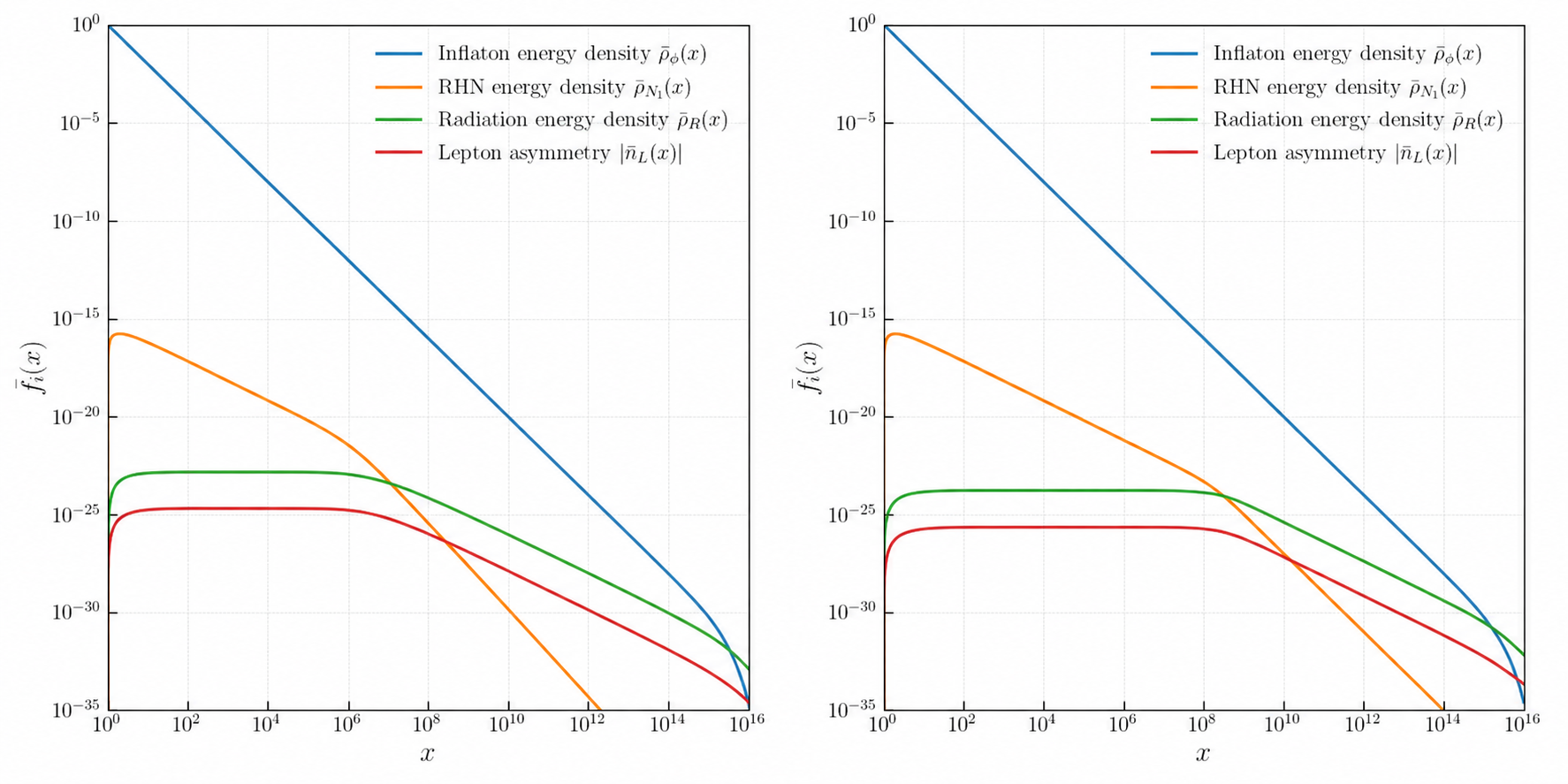}
    \caption{\it Representative evolution of the reheating and nonthermal leptogenesis functions $\bar{f}_i(x)$ supposing 
    $k_{\phi \to {N_1}{N_1}}\sim 10^{-15}$, $\varepsilon_1\sim - 10^{-5}$, $\bar{M}_{N_1}\sim 10^{-3}$ 
    and two reference (normalized) values of the RHN decay rate into massless Standard Model particles, 
    \emph{i.e.} $k_{{N_1} \to RR}\sim 10^{-6}$, corresponding to an effective or reference
    light neutrino mass $\tilde{m}_1\sim 10^{-3}\text{ eV} <m^{\text{max}}_{\nu}$ (left plot) 
    and $k_{{N_1} \to RR}\sim 10^{-8}$ corresponding to a very small effective mass, $\tilde{m}_1\sim 10^{-5}$ eV (right plot).
    Increasing the heavy-neutrino decay rate shortens its lifetime, leading to an earlier onset of exponential suppression and a reduced duration of the radiation energy density and lepton-asymmetry number density plateaus.
    It is also interesting to see that although $k_{\phi \to {N_1}{N_1}}/k_{{N_1} \to RR}\sim 10^{-10}$ for instance, 
    the \emph{plateau} regime is a nontrivial features of the dynamics. 
    The graceful exit towards the radiation dominance and the freezing of the lepton asymmetry always occur at $x\sim k_{\phi \to {N_1}{N_1}}^{-1}$.}
    \label{Fig:3}
\end{figure*}

In this setup, the inflaton 
is treated as a purely massive nonrelativistic degree of freedom
with a matter-like equation-of-state $w_\phi=0$. It is also assumed to be sufficiently heavy
to never reach chemical equilibrium with RHN or light Standard Model degrees of freedom. 
In addition, its decay is taken to proceed dominantly into RHNs (see Sec. \ref{Pseudoscalaron interactions with fermions}). 
The RHN $N_1$ responsible for both the production of SM particles and lepton asymmetry, is also assumed to be heavy, $M_{N_1} \lesssim m_\phi/2$, and thus produced non relativistically.  
In any case, possible transient relativistic stages would tend to be short-lived and to not (crucially)
affect the late-time dynamics relevant for reheating and asymmetry generation \cite{Plumacher:1996kc}.
In addition inverse decays and scatterings from the thermal bath into RHNs are neglected.
For more details on the Einstein-Boltzmann system and adopted conventions, see Appendix \ref{appB}.
It is convenient to reformulate the dynamics in dimensionless variables (see, \textit{e.g.}, \cite{DiMarco:2021xzk}) as
\begin{eqnarray}
\bar{\rho}_\phi'(x) + 3 H(x)\,\bar{\rho}_\phi(x)
&=& - \frac{2}{3} k_{\phi \to RR}\,\bar{\rho}_\phi(x)
     - \frac{2}{3} k_{\phi \to N_1 N_1}\,\bar{\rho}_\phi(x) ,\label{BEnorm1}\\
\bar{\rho}_{N_1}'(x) + 3 H(x)\,\bar{\rho}_{N_1}(x)
&=& + \frac{2}{3} k_{\phi \to {N_1}{N_1}}\,\bar{\rho}_\phi(x)
     - \frac{2}{3} k_{{N_1} \to RR}\,\bar{\rho}_{N_1}(x) ,\label{BEnorm2}\\
\bar{\rho}_R'(x) + 4 H(x)\,\bar{\rho}_R(x)
&=& + \frac{2}{3} k_{\phi \to RR}\,\bar{\rho}_\phi(x)
     + \frac{2}{3} k_{{N_1} \to RR}\,\bar{\rho}_{N_1}(x) ,\label{BEnorm3}\\
\bar{n}_{L}'(x) + 3 H(x)\,\bar{n}_{L}(x)
&=&  \frac{2}{3} \varepsilon_1 \, k_{{N_1} \to RR}\, \frac{\bar{\rho}_{N_1}(x)}{\bar{M}_{N_1}}    ,
\label{BEnorm4}\end{eqnarray}
where the evolution variable $x$ is defined as the cosmic time normalized to the characteristic time scale at the end of inflation,
\begin{eqnarray}
x = \frac{t}{t_{\rm end}}, \quad
t_{\text{end}}\sim \frac{2}{3H_{\text{end}}} .
\end{eqnarray}
The reheating quantities are normalized with appropriate powers of the inflaton energy density at the end of inflation,
\begin{eqnarray}
\bar{\rho}_\phi = \frac{\rho_\phi}{\rho_{\rm end}}, \qquad
\bar{\rho}_{N_1} = \frac{\rho_{N_1}}{\rho_{\rm end}}, \qquad
\bar{\rho}_R = \frac{\rho_R}{\rho_{\rm end}}, \qquad
\bar{n}_{L} = \frac{n_{L}}{\rho^{3/4}_{\rm end}}, \qquad
\bar{M}_{N_1} = \frac{M_{N_1}}{\rho^{1/4}_{\rm end}} 
\end{eqnarray}
and the normalized decay rates are defined as
\begin{equation}
k_{\phi \to RR} = \frac{\Gamma_{\phi \to RR}}{H_{\rm end}}, \qquad
k_{\phi \to {N_1}{N_1}} = \frac{\Gamma_{\phi \to {N_1}{N_1}}}{H_{\rm end}}, \qquad
k_{{N_1} \to RR} = \frac{\Gamma_{{N_1} \to RR}}{H_{\rm end}} ,
\end{equation}
with the total normalized inflaton decay rate given by 
\begin{equation}
k_\phi = k_{\phi \to RR} + k_{\phi \to {N_1}{N_1}} .
\end{equation}
The obvious initial conditions are then
\begin{equation}
\bar{\rho}_\phi(1) = 1, \qquad
\bar{\rho}_{N_1}(1) = 0, \qquad
\bar{\rho}_R(1) = 0, \qquad
\bar{n}_{L}(1) = 0 ,
\end{equation}
while the Hubble rate, in this parameterization, takes the form 
\begin{equation}
H^2(x)
= \left( \frac{2}{3} \right)^2
\left[
\bar{\rho}_\phi(x)
+ \bar{\rho}_{N_1}(x)
+ \bar{\rho}_R(x)
\right].
\end{equation}
The obtained system of coupled equations
can be solved in terms of the scale factor $a(x)$. Exploiting the standard definition of the Hubble rate, $H(x)= a'(x)/ a(x)$, one gets
\begin{eqnarray}
\bar{\rho}_\phi(x) &=& \left(\frac{a_{\rm end}}{a(x)}\right)^3 e^{-\frac{2}{3} k_\phi (x-1)}  ;\label{InfNorDens}\\[0.4cm]
\bar{\rho}_{N_1}(x) &=& \left(\frac{a_{\rm end}}{a(x)}\right)^3 \frac{k_{\phi \to {N_1}{N_1}}} {k_{{N_1} \to RR} 
\left( 1 - \frac{k_{\phi \to {N_1}{N_1}}}{k_{{N_1} \to RR}} \right)} 
\left[ e^{-\frac{2}{3} k_\phi (x-1)} - e^{-\frac{2}{3} k_{{N_1} \to RR} (x-1)} \right] ; \\[0.4cm]
\bar{\rho}_R(x) &=& \left(\frac{a_{\rm end}}{a(x)} \right)^4 \frac{2}{3} \frac{1}{\left(1 - \frac{k_\phi}{k_{{N_1} \to RR}}\right)}
\Bigg[k_\phi \left( 1 - \frac{k_{\phi \to RR}}{k_{{N_1} \to RR}} \right) I_\phi(x) - k_{\phi \to {N_1}{N_1}} \ I_{N_1}(x) \Bigg]  ; \\[0.4cm]
\bar{n}_{L}(x) &=& \left( \frac{a_{\rm end}}{a(x)} \right)^3 \frac{\text{BR}(\phi \to {N_1}{N_1})}{1 - \frac{k_\phi}{k_{{N_1} \to RR}}}
\frac{\varepsilon_1}{\bar{M}_{N_1}} \left\{ 1 - e^{-\frac{2}{3} k_\phi (x-1)} - \frac{k_\phi}{k_{{N_1} \to RR}} 
\left[ 1 - e^{-\frac{2}{3} k_{{N_1} \to RR} (x-1)} \right] \right\} ,
\end{eqnarray}
where the functions $I_\phi(x)$ and $I_{N_1}(x)$ entering the radiation solution are defined by
\begin{eqnarray}
I_\phi(x) &=& \int_1^x du\, \frac{a(u)}{a_{\rm end}} e^{-\frac{2}{3} k_\phi (u-1)},
\qquad
I_{N_1}(x) = \int_1^x du\, \frac{a(u)}{a_{\rm end}} e^{-\frac{2}{3} k_{{N_1} \to RR} (u-1)} .
\end{eqnarray}
In general, these functions do not admit closed explicit solutions in terms of the dimensionless time variable $x$.
Nevertheless, analytically controlled regimes can be identified. 
Indeed, as previously discussed, 
the inflaton dominates the expansion during reheating, \emph{namely}
\begin{eqnarray}
H^2(x)\simeq \left(\frac{2}{3}\right)^2\ \bar{\rho}_{\phi}(x) ,
\label{ApproxHub}\end{eqnarray}
and it decays slowly and exclusively into RHNs ensuring a dynamically negligible $k_{\phi \to RR}$ parameter
and an inflaton branching ratio into RHN of order unity:
\begin{eqnarray}
k_{\phi \to RR} \sim 0, \quad
k_{\phi} \sim k_{\phi \to {N_1}{N_1}}, \quad
\text{BR}(\phi \to {N_1}{N_1}) \sim 1 .
\end{eqnarray}
As a result, inserting the inflaton density solution of Eq.\eqref{InfNorDens} in Eq. \eqref{ApproxHub}, the evolution of the dimensionless scale factor comes out to be
\begin{eqnarray}
\frac{a(x)}{a_{\text{end}}}\simeq 
\Bigg[1 + \frac{2}{3k_{\phi \to {N_1}{N_1}}} - \frac{3}{2k_{\phi \to {N_1}{N_1}}}
e^{-\frac{k_{\phi \to {N_1}{N_1}}}{3}(x-1)}
\Bigg]^{2/3}
\sim x^{2/3},
\label{scalefactorbeh}\end{eqnarray}
neatly deriving by the fact that the quantity $k_{\phi \to {N_1}{N_1}} (x-1)$ is naturally small, 
being the ratio between the normalized cosmic time and the normalized inflaton lifetime.
Under these conditions, the above solutions can be written explicitly as functions of $x$, namely
\begin{eqnarray}
\bar{\rho}_\phi(x)
&=&
\frac{1}{x^2}
e^{-\frac{2}{3} k_{\phi \to {N_1}{N_1}} (x-1)}
 ; \label{BEsol1}\\[0.4cm]
\bar{\rho}_{N_1}(x)
&=&
\frac{1}{x^2}
\frac{k_{\phi \to {N_1}{N_1}}}
{k_{{N_1} \to RR}
\left(
1 - \frac{k_{\phi \to {N_1}{N_1}}}{k_{{N_1} \to RR}}
\right)}
\left[
e^{-\frac{2}{3} k_{\phi \to {N_1}{N_1}} (x-1)}
-
e^{-\frac{2}{3} k_{{N_1} \to RR} (x-1)}
\right]
 ; \label{BEsol2}\\[0.4cm]
\bar{\rho}_R(x)
&=&
\frac{2}{3x^{8/3}}
\frac{k_{\phi \to {N_1}{N_1}}}{\left(1 - \frac{k_{\phi \to {N_1}{N_1}}}{k_{{N_1} \to RR}}\right)}
\Bigg[ I_\phi(x) - I_{N_1}(x) \Bigg] \label{integralphiN}
 ; \label{BEsol3}\\[0.4cm]
\bar{n}_{L}(x)
&=&
\frac{1}{x^2}
\frac{1}{1 - \frac{k_{\phi \to {N_1}{N_1}}}{k_{{N_1} \to RR}}}
\frac{\varepsilon_1}{\bar{M}_{N_1}}
\left\{
1
-
e^{-\frac{2}{3} k_{\phi \to {N_1}{N_1}} (x-1)}
-
\frac{k_{\phi \to {N_1}{N_1}}}{k_{{N_1} \to RR}}
\left[
1
-
e^{-\frac{2}{3} k_{{N_1} \to RR} (x-1)}
\right]
\right\}.
\label{BEsol4}\end{eqnarray}
It should be noticed that, using the expression in Eq. \eqref{scalefactorbeh}, the integrals in Eq. \eqref{integralphiN} assume a compact form 
in terms of the normalized lower incomplete Gamma function $P$,
defined as \cite{Abramowitz:1965zz}
\begin{equation}
    \Gamma(a)  P(a,x) = \int_0^x  t^{a -1} \ e^{-t} \ dt .
\end{equation}
Indeed, one has 
\begin{equation}
I_f(x) = A_f^{-\alpha} e^{A_f} \Big[ \Gamma(\alpha) \,
P(\alpha, A_f \, x) - \Gamma(\alpha) \, P(\alpha,A_f) \Big],
\end{equation}
where $\alpha = 5/3$ and 
\begin{eqnarray}
f =
\left\{
\begin{array}{ll}
\phi, & A_\phi = \dfrac{2 k_{\phi \to {N_1}{N_1}}}{3} \\[0.2cm]
{N_1},    & A_{N_1}    = \dfrac{2 k_{{N_1} \to RR}}{3}
\end{array}
\right. 
\end{eqnarray}
The normalized system of Boltzmann Eqs. \eqref{BEnorm1}-\eqref{BEnorm4} has been numerically integrated via
the backward differentiation formula (BDF) method and supposing $|\gamma|\sim 10^{-2}$, with
a couple of reference values of the ${N_1}$ decay rates, see Fig. \ref{Fig:3}.
The analytical solutions provided in Eqs. \eqref{BEsol1}-\eqref{BEsol4} concretely match such numerical results 
suggesting the robustness of the approximations.
In general, the overall dynamics is initially controlled, as expected, by the inflaton component,
whose slow decay settles the timescale of the energy transfer. 
Larger RHN decay rates lead to a shorter and more pronounced intermediate plateau, 
reflecting a more efficient and rapid conversion of the nonthermal ${N_1}$ population into radiation and $B-L$ asymmetry. 
Conversely, smaller decay rates prolong the duration of the plateau 
and delay the onset of the radiation domination over the RHN sector. 
At sufficiently late times, however, the solutions approach a similar asymptotic behavior, 
as the inflaton energy density becomes negligible and the system evolves toward radiation domination.

In the extreme limit in which the RHN decay is effectively instantaneous compared to the inflaton decay, \textit{i.e.} 
for $k_{\phi \to {N_1}{N_1}}/k_{{N_1} \to RR}\ll 10^{-10}$, one obtains
\begin{eqnarray}
\bar{\rho}_\phi(x)
&\sim&
\frac{1}{x^2}
e^{-\frac{2}{3} k_{\phi \to {N_1}{N_1}} (x-1)}
\\[0.4cm]
\bar{\rho}_{N_1}(x)
&\sim&
\frac{1}{x^2}
\frac{k_{\phi \to {N_1}{N_1}}}
{k_{{N_1} \to RR}}
e^{-\frac{2}{3}k_{\phi \to {N_1}{N_1}}(x-1)}
\left[ 1-
e^{-\frac{2}{3} k_{{N_1} \to RR} (x-1)}
\right]
\\[0.4cm]
\bar{\rho}_R(x)
&\sim&
\frac{2k_{\phi \to {N_1}{N_1}}}{5x} \left( 1 - \frac{1}{x^{5/3}} \right) \label{leadendens}
\\[0.4cm]
\bar{n}_{L}(x)
&=&
\frac{1}{x^2}
\frac{\varepsilon_1}{\bar{M}_{N_1}}
\Bigg[ 1 - e^{-\frac{2}{3}k_{\phi \to {N_1}{N_1}}(x-1)}  \Bigg].
\label{leadleptasim}
\end{eqnarray}

In principle, if the RHN decay width is not completely negligible compared to the inflaton decay width, 
these expressions fail to reproduce the early-time transient regime captured by a numerical integration of the full system. Nevertheless, they accurately describe the late-time evolution for $x$ approaching the reheating time scale, $x_{\text{reh}}\sim k_{\phi \to {N_1}{N_1}}^{-1}$, and therefore provide reliable estimates of the asymmetry prior to the onset of the standard radiation dominance. 
In particular, the comparison of the leading-order expression for the radiation energy density Eq. \eqref{leadendens} around $x\sim x_{\text{reh}}$ 
with the standard definition of energy density for a gas of relativistic particles $\rho_R(x_{\rm reh}) = \pi^2 g_{E}(x_{\rm reh})T_{\rm reh}^4/30$,
yields the reheating temperature
\begin{eqnarray}
T_{\text{reh}}\sim \left( \frac{180}{5\pi^2g_{\text{E}}(x_{\text{reh}})} \right)^{1/4}\sqrt{M_p \Gamma_{\phi \to {N_1}{N_1}}} .
\end{eqnarray}
where $g_{E}$ represents the number of (energy) relativistic degrees of freedom
This result allows to parameterize the leading-order expression for the lepton asymmetry of Eq. \eqref{leadleptasim} 
normalized to the entropy density, that can be written as
\begin{eqnarray}
\frac{n_{L}}{s}\sim \frac{3}{2}\,\varepsilon_1\,\frac{T_{\text{reh}}}{M_{N_1}}
\end{eqnarray}
and finally becomes, for $M_{N_1} \lesssim m_{\phi}/2$,
\begin{eqnarray}
\frac{n_{L}}{s}\sim 3\,\varepsilon_1\,\frac{T_{\text{reh}}}{m_\phi}.
\end{eqnarray}
The resulting electroweak-induced baryon asymmetry is then computed as
\begin{equation}
    \frac{n_B}{s} = - c_{\text{sph}} \frac{n_{L}}{s} ,
\label{bauonentro}\end{equation}
where $c_{\text{sph}}$ is the corresponding sphaleron conversion coefficient given by
\begin{equation}
    c_{sph}=\frac{8 n_F + 4 n_S}{22 n_F + 13 n_S}, 
\label{sphalconv}\end{equation}
with $n_F$ the number of fermion families and $n_S$ the number of Higgs doublets.
In the non-supersymmetric case where only SM matter is present in the matter sector
of the complete model, $n_F=3$ and $n_S=1$, so $c_{sph} = 28/79\sim 0.35$.
The Einstein-Cartan scenario with $\gamma\sim-1/100$ is strongly compatible with CMB data if the
observed cosmological perturbations are produced 
when $N_e\sim 60$ $e$-folds before the end of inflation 
(as seen in the previous sections). 
This setup typically predicts an inflaton mass of order $10^{14}$ GeV. 
A RHN mass order $M_{N_1}\sim m_{\phi}/2 \sim  5\times10^{13}$ GeV provides $\Gamma_{\phi \to {N_1}{N_1}}\sim 1.6 \times10^{-1}$ GeV, 
$\Gamma_{{N_1}\to RR}\sim 1.6 \times 10^{9}$ GeV
(with $\tilde{m}_1\sim 10^{-3}$ eV) and a reheating temperature $T_{\text{reh}} \lesssim 3 \times10^8$ GeV. 
The Davidson-Ibarra bound for the maximum value of the CP asymmetry parameter is $|\varepsilon_1|\sim 3.4\times 10^{-2}$. 
The case $|\varepsilon_1|\sim 3\times 10^{-5}$ -- also used in the numerical computations --
is well within the Davidson-Ibarra bound and therefore compatible with the latest neutrino measurements on the sum of masses
and can easily accommodate for a final baryon asymmetry in the range $n_B/s\sim 10^{-10}-10^{-11}$, consistent with the observed value.
Specific numerical results, obtained via Eq. \eqref{bauonentro}, are listed in Table~\ref{tab:2}.

\begin{table}[t]
\centering
\small
\setlength{\tabcolsep}{18pt}
\renewcommand{\arraystretch}{1.1}

\begin{tabular}{p{0.10\textwidth} c c c c c}
\hline\hline

\mbox{Barbero--Immirzi $\gamma$}
& $H_{\text{end}}$ (GeV) 
& $m_\phi$ (GeV) 
& $\Gamma_{\phi \rightarrow N_1N_1}$ (GeV) 
& $T_{\text{reh}}$ (GeV) 
& $n_B/s$ \\
\hline

$-1/150$ & $4.30\times10^{13}$ & $1.61\times10^{14}$ & $5.21\times10^{-2}$ & $1.53\times10^{8}$ & $5.01\times10^{-11}$ \\
$-1/140$ & $4.47\times10^{13}$ & $1.67\times10^{14}$ & $6.36\times10^{-2}$ & $1.69\times10^{8}$ & $5.54\times10^{-11}$ \\
$-1/130$ & $4.66\times10^{13}$ & $1.74\times10^{14}$ & $7.84\times10^{-2}$ & $1.88\times10^{8}$ & $6.15\times10^{-11}$ \\
$-1/120$ & $4.85\times10^{13}$ & $1.81\times10^{14}$ & $9.77\times10^{-2}$ & $2.10\times10^{8}$ & $6.87\times10^{-11}$ \\
$-1/110$ & $5.06\times10^{13}$ & $1.89\times10^{14}$ & $1.23\times10^{-1}$ & $2.36\times10^{8}$ & $7.72\times10^{-11}$ \\
$-1/100$ & $5.28\times10^{13}$ & $1.97\times10^{14}$ & $1.58\times10^{-1}$ & $2.67\times10^{8}$ & $8.74\times10^{-11}$ \\
$-1/90$ & $5.50\times10^{13}$ & $2.05\times10^{14}$ & $2.06\times10^{-1}$ & $3.05\times10^{8}$ & $9.98\times10^{-11}$ \\
$-1/80$ & $5.73\times10^{13}$ & $2.14\times10^{14}$ & $2.75\times10^{-1}$ & $3.52\times10^{8}$ & $1.15\times10^{-10}$ \\

\hline\hline
\end{tabular}

\caption{Estimates of the main postinflationary quantities as function of the Barbero--Immirzi parameter $\gamma$. The quantities are computed by using $p=2$, assuming the observed cosmological perturbations are stretched outside the Hubble horizon at $N_e \sim 60$ $e$-folds before the end of inflation, a neutrino mass $M_{N_1} \sim 5\times 10^{13}$ GeV and a CP violation parameter $|\epsilon_1|\sim 3\times 10^{-5}$, consistent with the Davidson-Ibarra bound.}
\label{tab:2}
\end{table}

\section{Conclusions and Prospects}
\label{Conclusions and Prospects}

The Einstein-Cartan-Holst pseudoscalaron models proposed in \cite{Pradisi:2022nmh,DiMarco:2023ncs} describe GR together with a single pseudoscalar inflaton field dual, via a non-linear term in the Holst curvature, to the (pseudo)vector components of the dynamical contortion. The pseudoscalar inflaton is subject to a potential that  can drive a successful single-field slow-roll inflationary phase, followed by a reheating epoch dominated by the corresponding vacuum modes. 
In this paper, the inflation sector has been coupled to a type I seesaw extension of the SM containing three sterile hierarchical Majorana right-handed neutrinos, with the lightest and cosmologically relevant state having a bare mass below $10^{14}$ GeV. 
All matter fermions (both in the SM and in the right-handed neutrino sector) have been assumed to be minimally coupled to gravity.

A central result of the analysis is that the inflaton gets a \emph{universal coupling} to matter fermions, dictated by the contortion part of the connection. Since the corresponding decay rates are proportional to the squared masses of the final-state fermions, 
the inflaton can efficiently decay only into the lightest Majorana right-handed neutrino $N_1$.
The nonthermal generated RHN particles then trigger a lepton asymmetry production, subsequently converted into a baryon asymmetry
via the standard electroweak sphaleron processes.
It has been shown that, for a reasonable range of the Barbero-Immirzi parameter
$\gamma \in [-1/80,-1/150]$, and for suitable values of the lightest right-handed neutrino mass, the resulting cosmological history
simultaneously accounts for a viable single-field slow-roll inflationary stage
and a nonthermal leptogenesis (and baryogenesis) mechanism,
fully compatible with the current constraints from CMB, BAO and BBN data. 
An important phenomenological feature of this scenario is that the reheating temperature is quite lower than in the
thermal case, paving the way to a sensible coupling to supergravity and thus to its ultraviolet completions, (super)string
or M-theory, avoiding the gravitino problem. These directions will be pursued in forthcoming papers.

It would be interesting to promote this class of models to a portal for inflation and leptogenesis, trying to include dark matter as well, and to get a successful explanation for the dark energy of the Universe. The discussed models, as shown, are based on a hypothetical and peculiar gravitational/inflationary sector coupled to seesaw extensions of the SM that account for neutrino physics and allow for matter generation. Although many parameters related to neutrino physics have been precisely measured, seesaw scenarios, despite their theoretical appeal, remain poorly constrained because of their large parameter space. As a consequence, it is very hard to provide precise and stringent predictions on some important observables like the Majorana RHN number, the fermionic nature of active neutrinos and their mass hierarchy, together with an understanding of the structure of PMNS matrix and of the leptonic Yukawa couplings.
The relation of the gravity sector with the so called \emph{swampland conjectures} would be as well worth of investigations. 
Some of the mentioned issues will be addressed in the next generation of planned proper experiments.

\section{Acknowledgments}
The authors thank M. Migliaccio for suggestions about experimental data treatments and for illuminating discussions.
A.D.M. has been supported by the G4S$\_$2.0 project, developed under the auspices of the Italian Space Agency (ASI) within the frame of the Bando
Premiale CI-COT-2018–085 with co-participation of the
Italian Institute for Astrophysics (INAF) and the Politecnico di Torino (POLITO).

\appendix
\section{Fermionic Conventions}
\label{appA}
In this Appendix the adopted four-dimensional spinor conventions and their relation to Einstein-Cartan geometry are summarized.  In order to couple spinors to gravity one needs a spin manifold, namely a manifold that admits a globally defined non-holonomic orthonormal basis of the tangent space, $e_a$. It is related to the "coordinate" basis via the vierbein or tetrad, defined by the relation
\begin{equation}
    e^a_{~\mu} e^b_{~\nu}  \, \eta_{ab}  = g_{\mu\nu} ,
\end{equation}
where $e_a = e_a^{~\mu} \, \partial_{\mu}$. It also follows that $e=\det(e^a_{~\mu}) = \sqrt{- g}$. Tetrad is a sort of square root of the metric and can be used to transform curved to flat indices. If the tetrad exists, it is possible to choose a flat metric in any point of the tangent space. It means that the theory has a gauge symmetry that coincides with local Lorentz transformations, since point by point in the base manifold it is possible to choose arbitrary orthonormal basis of the tangent space that differ among themselves by (flat) Lorentz transformations. One may also introduce a (spin) connection, namely a one-form connection $\omega^{a}_{~ \ b \mu}$ taking values in the adjoint representation of the Lorentz group ({\it i.e.} with the property $\omega^{ab}_{~ \ ~ \mu} = - \omega^{ba}_{~ \ ~ \mu}$).  The corresponding covariant derivative acts as usual on (flat) vectors. For instance, the Lorentz metric is automatically covariantly constant, $\mathcal{D}_{\mu} \eta_{ab} = 0$, rendering the manifold metric (of Einstein-Cartan type).
The covariant derivative on fermions is the usual one for a principal bundle with gauge group $SO(3,1)$, namely
\begin{equation}
    \mathcal{D}_{\mu} \psi = \partial_{\mu} \psi + \frac{1}{4} \omega^{ab}_{~ \ ~ \mu} \, \gamma_{ab} \, \psi
\end{equation}
where $\gamma_{ab}$ are the generators of the Lorentz group on spinors in flat space, 
\begin{equation}
    \gamma^{ab} = \frac{1}{2} [\gamma^a, \gamma^b] .
\end{equation}
In a first order (Cartan) description of differential calculus, the properties of an Einstein-Cartan manifold are contained in the so called structure equations. The first one connects torsion and vierbein,
\begin{equation}
    T^a_{ ~\ \mu\nu}= \mathcal{D}_{\nu} e^{a}_{~ \mu} - \mathcal{D}_{\mu} e^{a}_{~ \nu} ,
\end{equation}
while the second relates curvature and connection
\begin{equation}
    {\mathcal R}_{~\ ~\mu\nu}^{a b} = \partial_{\mu} \omega^{ab}_{~\ ~\ \nu} + \omega^{a}_{~\ c \mu} \omega^{cb}_{~\ ~\ \nu} - (\mu \leftrightarrow \nu) .
\end{equation}
Spinor actions on curved manifolds can be constructed by using the usual "minimal prescription". In particular, for a Dirac (or Majorana) fermion, the manifestly hermitean form is 
\begin{equation}
     \mathcal{S}_{f} = \int d^4x \, e \, \zeta \, \left[ \frac{i}{2} \, \left(\overline{\psi} \, \gamma^{\mu}\,   \mathcal{D}_{\mu} \, \psi - \overline{ \mathcal{D}_{\mu} \psi} \ \gamma^{\mu} \, \psi \right) - m \overline{\psi} \psi \right] \, ,
\end{equation}
where $\zeta$ is $1$ ($1/2$) for Dirac (Majorana) spinors and  $\gamma^{\mu} = e^{\mu}_{~a} \gamma^{a}$.  It is useful to recall that Majorana spinors satisfy $\psi=\psi^c$, where $\psi^c = C \bar{\psi}^T$, with $C$ unitary such that $C \gamma^a C^{-1} = - \gamma^{a T}$.

Gravitational Lagrangians in the first order formulations are  obtained in a similar way. The Einstein-Hilbert action results
\begin{equation}
     \mathcal{S}_{\text{EH}} = - \frac{M_P^2}{2} \, \int d^4x \, e \, e^{\mu}_{~a} \, e^{\nu}_{~b} \, {\mathcal R}_{~\ ~\mu\nu}^{a b}(\omega) ,
\end{equation}
where it is easy to recognize the scalar curvature ${\mathcal R}$, and the dependence of the curvature tensor by the spin connection has been stressed. Finally, the so-called Holst term can be written as
\begin{equation}
     \mathcal{S}_{\text{Holst}} = \frac{M_P^2}{4 \gamma} \, \int d^4x \, e \, e^{\mu}_{~a} \, e^{\nu}_{~b} \, \varepsilon^{ab}_{~\ ~\ cd} \, {\mathcal R}_{~\ ~\mu\nu}^{c d}(\omega) ,
\end{equation}
where $\gamma$ is the Barbero-Immirzi parameter and the Holst curvature ${\mathcal R'}$ has been written in the flat basis.

\section{Parameters for nonthermal leptogenesis}
\label{appB}

A general Einstein-Boltzmann system for reheating and type-I seesaw nonthermal leptogenesis is expected to describe the coupled evolution of the inflaton number density $n_{\phi}$, the lightest right-handed neutrino number density $n_{{N_1}}$, the radiation energy density $\rho_{R}$ and the lepton asymmetry, defined as the difference between the number densities of leptons and antileptons,
\begin{equation}
n_L = n_\ell - n_{\bar\ell}.
\end{equation}
In a schematic form, one may write 
\begin{eqnarray}
\dot{n}_{\phi}+3H\,n_{\phi} &= S_{\phi},\\[4pt]
\dot{n}_{N_1}+3H\,n_{N_1} &= S_{N_1},\\[4pt]
\dot{\rho}_{R}+4H\,\rho_{R} &= S_{R},\\[4pt]
\dot{n}_{L}+3H\,n_{L} &= S_{L},
\label{nonthevol}
\end{eqnarray}
that must be accompanied by a proper expression for the Hubble rate 
and a reliable choice for the (numerical densities and radiation energy density) initial conditions.
Here, the evolution is conveniently formulated in terms of cosmic time or, equivalently, in terms of comoving quantities. 
A description in terms of dimensionless variable like
\begin{equation}
z = \frac{M_{N_1}}{T},
\end{equation}
commonly used in standard thermal leptogenesis, is indeed impossible. 
The reason is that nonthermal leptogenesis involves the simultaneous evolution of 
inflaton, radiation and right-handed neutrinos during reheating, 
a phase where the temperature does not generically result in a monotonic variable of the dynamics, 
differently from the cases of a pure radiation-dominated epoch or of a phase where radiation is fully decoupled from the other components. 
The source terms \(S_X\) in Eq. \eqref{nonthevol}, with \(X=\phi,{N_1},R,L\), 
encode the microscopic particle-physics processes governing the dynamics of the different quantities, \textit{i.e.} their production and dilution. 
In general, these source terms receive contributions both from thermal equilibrium and out-of-equilibrium processes, 
depending on the interactions in the underlying model. 
A representative set of equations with explicit source terms generically takes the following form:
\begin{eqnarray}
\dot{n}_{\phi}+3H\,n_{\phi} &=& 
-\Gamma_{\phi}\,n_{\phi}, \label{bews1}\\[4pt]
\dot{n}_{N_1}+3H\,n_{N_1} &=& 
-\Gamma^{\mathrm{th}}_{{N_1}\to RR}\left(n_{N_1}-n_{N_1}^{\mathrm{eq}}\right)
+2\Gamma_{\phi\to {N_1}{N_1}}\,n_{\phi} + {\mathcal{S}}_{{N_1}}, 
\label{bews2}\\[4pt]
\dot{\rho}_{R}+4H\,\rho_{R} &=&
\Gamma_{\phi\to RR}\,m_{\phi}n_{\phi}
+E_{N_1}\Gamma^{\mathrm{th}}_{{N_1}\to RR}\left(n_{N_1}-n_{N_1}^{\mathrm{eq}}\right),
\label{bews3}\\[4pt]
\dot{n}_{L}+3H\,n_{L} &=&
\varepsilon_1\,\Gamma^{\mathrm{th}}_{{N_1}\to RR}\left(n_{N_1}-n_{N_1}^{\mathrm{eq}}\right)
-\Gamma_{w}\,n_L. \label{bews4}
\end{eqnarray}
The corresponding Hubble rate reads
\begin{eqnarray}
H^{2}(t) = \frac{1}{3M^2_p}\left[
\rho_{\phi}(t) + \rho_{N_1}(t) + \rho_R(t)
\right] ,
\end{eqnarray}
where the (non relativistic) inflaton energy density can be written as $\rho_\phi = m_\phi \, n_\phi$ and 
the heavy RHN neutrino energy density can be parameterized by a reference total neutrino energy $E_{N_1}$ as
\begin{eqnarray}
    \rho_{N_1} = E_{N_1}\,n_{N_1},\quad \text{with} \quad E_{N_1} = \sqrt{M^2_{N_1} + p^2_{N_1}} \, ,
\end{eqnarray}
where $p_{N_1}$ is an effective RHN momentum, obtained by red-shifting the one produced by the inflaton decay kinematics
\begin{eqnarray}
    p_{N_1} = p_{N_1}(t_{\text{end}})\,\frac{a}{a_{end}}, \quad 
   \text{with} \quad p_{N_1}(t_{\text{end}}) = \frac{m_\phi}{2}\sqrt{1-\frac{4M^2_{N_1}}{m^2_\phi}}.
\end{eqnarray}
Finally, the initial condition of the problem can be safely tuned as
\begin{eqnarray}
n_{\phi}(t_{\text{end}}) = n_\phi(\phi_{\text{end}}), \quad
n_{N_1}(t_{\text{end}}) \simeq 0, \quad
\rho_R(t_{\text{end}}) \simeq 0, \quad
n_{L}(t_{\text{end}}) \simeq 0.
\end{eqnarray}

It is useful to analyze in detail the various source contributions on the right hand side of the Boltzmann system.
\(w_{\phi}\) is the inflaton equation-of-state parameter, which accounts for deviations from the purely matter-like behavior of coherent inflaton oscillations. \(\varepsilon_1\) measures the \(CP\)-asymmetry produced in leptonic ${N_1}$ decays. \(\Gamma_{\phi}\) denotes the total inflaton 
decay rate, typically given by the sum of the two-body decay rates into radiation
($\phi \to RR$)
and  RHNs 
($\phi \to {N_1}{N_1}$): \(\Gamma_{\phi}=\Gamma_{\phi\to RR}+\Gamma_{\phi\to {N_1}{N_1}}\). Its explicit form depends, of course,  on the inflaton nature and couplings. \(\Gamma^{\mathrm{th}}_{{N_1}\to RR}\) is the ``thermally averaged'' decay rate of the right-handed neutrino into radiation, consisting basically of SM Higgs particles and leptons, \textit{i.e.} $({N_1} \to \ell h)$, $({N_1} \to \bar \ell \bar h)$. It can be obtained as
\begin{equation}
\Gamma^{\mathrm{th}}_{{N_1}\to RR}
=
\Gamma_{{N_1}\to RR}\,
\frac{K_1(M_{N_1}/T)}{K_2(M_{N_1}/T)},
\end{equation}
where \(\Gamma_{{N_1}\to RR}\) is the corresponding zero-temperature RHN decay rate, with \(K_1\) and \(K_2\)  modified Bessel functions of the second kind \cite{Luty:1992un,Plumacher:1996kc,Barbieri:1999ma,Giudice:2003jh}. At late times, where $M_{N_1}>T$ is achieved, the thermally averaged decay width reduces to the zero-temperature one $\Gamma_{{N_1}\to RR}$. The equilibrium number density for a particle species $j$ is indicated by $n_j^{\mathrm{eq}}$ 
which, in the Maxwell-Boltzmann approximation, reads 
\begin{equation}
n_j^{\mathrm{eq}} = g_j\,\frac{m_j^2\,T}{2\pi^2}\, K_2\!\left(\frac{m_j}{T}\right) ,
\end{equation}
where $g_j$ represents the internal degrees of freedom of the species $j$ itself.
The term ${\mathcal{S}_{N_1}}$ schematically accounts for RHN number-changing scattering processes, including both RHN production and loss channels.
These may include reaction densities associated with processes such as $\ell\ell \to {N_1}{N_1}$, $hh \to {N_1}{N_1}$, 
as well as lepton-number-violating processes with $\Delta L = 1$ 
scatterings involving RHNs, leptons, quarks, and gauge bosons, depending on the interactions present in the model.
Finally, \(\Gamma_w\) represents the total washout term for the lepton asymmetry, \textit{i.e.} the set of processes that tend to suppress the generated asymmetry and to restore equal lepton and antilepton abundances in the plasma. Schematically, it is given by
\begin{equation}
\Gamma_w = W_{ID} + \Gamma_{\Delta L =1} + \Gamma_{\Delta L = 2},
\end{equation}
where
\begin{equation}
W_{ID}
=
\frac{\Gamma_{ID}}{2}
=
\Gamma^{\mathrm{th}}_{{N_1}\to RR}\,\frac{n_{N_1}^{\mathrm{eq}}}{2\,n_\ell^{\mathrm{eq}}}
\end{equation}
is the inverse-decay washout term, typically the dominant contribution, with \(n_\ell^{\mathrm{eq}}\) the equilibrium number density of leptons in the thermal bath.
The term $\Gamma_{\Delta L =1}$ represents the washout counterpart of the $\Delta L = 1$ RHN scattering processes mentioned above. 
The term $\Gamma_{\Delta L =2}$, instead, describes lepton-number-violating processes $2\to2$ with $\Delta L = 2$ scatterings such as 
$\ell h \leftrightarrow \bar{\ell}\bar{h}$, mediated by right-handed neutrinos.
It should be stressed that only the \emph{off-shell} or virtual intermediate RHN-state contributions should be included in this term, since the processes involving real intermediate RHN-states are already accounted for by the (sequence of) 
inverse decays ($\ell h \to {N_1}$ or $\bar \ell \bar h \to {N_1}$)
and decays (${N_1} \to \ell h$ or ${N_1}\to \bar\ell \bar h$), and 
must be subtracted to avoid double counting \cite{Kolb:1979qa}.

The system of Eqs. \eqref{bews1}-\eqref{bews4} contains the following information: the inflaton sector, which never experiences a thermal equilibrium phase
with the other components, is progressively emptied by its decays into radiation and right-handed neutrinos. 
The right-handed neutrino number density is sourced both by inflaton decays and by scattering processes, 
while it is reduced by its decays into radiation.
In general, the RHN can experience an (almost) relativistic phase followed by a nonrelativistic one, 
or only the latter if $M_{N_1}$ is very large.
The radiation energy density is fed both by inflaton and  (thermally averaged and vacuum) RHN decays. 
Finally, the lepton asymmetry is generated by the CP-violating processes measured by the $\varepsilon_1$ parameter, 
and simultaneously washed out by the previously mentioned inverse decay and scatterings.

In the limit of a purely non-relativistic massive inflaton and an almost heavy RHN with $M_{N_1}\lesssim m_{\phi}/2$
with weak interactions with the SM particles, 
it is reasonable to neglect the thermal equilibrium phase together with the related processes (thermal averaged decays and scatterings). 
Resorting also to a complete energy density-based formalism with $\rho_{\phi} = m_{\phi} n_{\phi}$ and $\rho_{N_1}\sim M_{N_1}n_{N_1}$, a simplified version of the system can be written in the form
\begin{eqnarray}
\dot{\rho}_\phi(t) + 3 H(t)\,\rho_\phi(t)
&=& - \Gamma_{\phi \to RR}\,\rho_{\phi}(t) - \Gamma_{\phi \to {N_1}{N_1}}\,\rho_{\phi}(t) , \\
\dot{\rho}_{N_1}(t) + 3 H(t)\,\rho_{N_1}(t)
&=& \Gamma_{\phi\to {N_1}{N_1}}\,\rho_\phi(t) - \Gamma_{{N_1}\to RR}\,\rho_{N_1}(t) ,\\
\dot{\rho}_R(t) + 4 H(t)\,\rho_R(t)
&=& \Gamma_{\phi\to RR}\,\rho_{\phi}(t) +\Gamma_{{N_1}\to RR}\,\rho_{N_1}(t) , \\
\dot{n}_{L}(t) + 3 H(t)\,n_{L}(t)
&=& \varepsilon_1 \,\Gamma_{{N_1}\to RR}\ \frac{\rho_{N_1}}{M_{N_1}}\label{enbml}.
\end{eqnarray}
The Hubble rate gets the same expression
\begin{eqnarray}
H^{2}(t) = \frac{1}{3M^2_p}\left[
\rho_{\phi}(t) + \rho_{N_1}(t) + \rho_R(t)
\right] ,
\end{eqnarray}
but now with $\rho_{\phi} = m_{\phi}n_{\phi}$, $\rho_{N_1} = M_{N_1} n_{N_1}$ 
and the proper initial conditions reduced to
\begin{eqnarray}
\rho_{\phi}(t_{\text{end}}) = \rho(\phi_{\text{end}}), \quad
\rho_{N_1}(t_{\text{end}}) \simeq 0, \quad
\rho_R(t_{\text{end}}) \simeq 0, \quad
n_{L}(t_{\text{end}}) \simeq 0 .
\end{eqnarray}
Some comments are in order concerning the sign convention adopted for the $CP$ asymmetry parameter $\varepsilon_1$ entering the Boltzmann system.
In the standard leptogenesis convention (used in the present manuscript), the $CP$ asymmetry is defined as\footnote{The definition holds both in the hypothetical initial equilibrium stage as well as in the following non equilibrium phase.}
\begin{equation}
\varepsilon_1=
\frac{\Gamma({N_1}\to \ell h)-\Gamma({N_1}\to \overline{\ell h})}
{\Gamma({N_1}\to \ell h)+\Gamma({N_1}\to \overline{\ell h})} \, ,
\end{equation}
so that a positive $\varepsilon_1$ corresponds to an excess of leptons over
antileptons, while a negative $\varepsilon_1$ corresponds to an excess of
antileptons.
It should be stressed that it is customary to indicate the \emph{baryon-lepton asymmetry} in terms of the quantity 
\begin{equation}
n_{B-L} = n_B-n_L \, ,
\label{bmlasymm}\end{equation}
where $n_B = n_b-n_{\bar{b}}$. The reason is the following. The sphalerons are anomalous $B+L$ violating processes. When a $B-L $ asymmetry is generated, it is converted into a $B$ asymmetry according to the relation
\begin{equation}
B = c_{sph} \ (B-L) 
\end{equation}
(with $c_{sph}$ in Eq. \eqref{sphalconv}), even though the asymmetry has been produced in a previous not equilibrium epoch. 
In the case of a completely negligible initial baryon asymmetry, \textit{i.e.}, $n_B = 0$
-- as customary in certain postinflationary scenarios --
the equation for the lepton asymmetry evolution of the Boltzmann system
could be conveniently substituted by the equivalent equation
\begin{equation}
    \dot{n}_{B-L}(t) + 3 H(t)\,n_{B-L}(t)
= - \varepsilon_1 \ \Gamma_{{N_1}\to RR}\ \frac{\rho_{N_1}}{M_{N_1}} ,
\end{equation}
where the minus sign for the source term comes from the identity in Eq. \eqref{bmlasymm}. 
In this paper, however, the lepton asymmetry is treated keeping the $n_L$ evolution.

\nocite{*}

\newcommand{\journal}[2]{\href{http://dx.doi.org/#1}{#2}}
\newcommand{\arxiv}[2]{\href{http://arxiv.org/abs/#1}{[arxiv:#1 #2]}}


\begin{thebibliography}{150}


\bibitem{Starobinsky:1980te}
A.~A.~Starobinsky,
``A New Type of Isotropic Cosmological Models Without Singularity,''
\journal{10.1016/0370-2693(80)90670-X}{Phys. Lett. B \textbf{91}, 99-102 (1980)}.


\bibitem{Guth:1980zm}
A.~H.~Guth,
``The Inflationary Universe: A Possible Solution to the Horizon and Flatness Problems,''
\journal{10.1103/PhysRevD.23.347}{Phys.\ Rev.\ D {\bf 23} (1981) 347}.


\bibitem{Linde:1981mu}
A.~D.~Linde,
``A New Inflationary Universe Scenario: A Possible Solution of the Horizon, Flatness, Homogeneity, Isotropy and Primordial Monopole Problems,''
\journal{10.1016/0370-2693(82)91219-9}{Phys.\ Lett.\  {\bf 108B} (1982) 389}.

\bibitem{Albrecht:1982wi}
A.~Albrecht and P.~J.~Steinhardt,
``Cosmology for Grand Unified Theories with Radiatively Induced Symmetry Breaking,''
\journal{10.1103/PhysRevLett.48.1220}{Phys.\ Rev.\ Lett.\  {\bf 48} (1982) 1220}.

\bibitem{Hawking:1981fz}
S.~W.~Hawking and I.~G.~Moss,
``Supercooled Phase Transitions in the Very Early Universe,''
\journal{doi:10.1016/0370-2693(82)90946-7}{Phys.\ Lett.\  {\bf 110B} (1982) 35}.

\bibitem{Linde:1983gd}
A.~D.~Linde,
``Chaotic Inflation,''
\journal{10.1016/0370-2693(83)90837-7}{Phys.\ Lett.\  {\bf 129B} (1983) 177}.


\bibitem{Linde:1990flp}
A.~D.~Linde,
``Particle physics and inflationary cosmology,''
\journal{}{Contemp. Concepts Phys. \textbf{5}, 1-362 (1990)}
\arxiv{hep-th/0503203}{[hep-th]}.

\bibitem{Linde:2007fr}
A.~D.~Linde,
``Inflationary Cosmology,''
\journal{10.1007/978-3-540-74353-8\_1}{Lect. Notes Phys. \textbf{738}, 1-54 (2008)}
\arxiv{0705.0164}{[hep-th]}.

\bibitem{Olive:1989nu} 
K.~A.~Olive,
``Inflation,''
\journal{10.1016/0370-1573(90)90144-Q}{Phys.\ Rept.\  {\bf 190}, 307 (1990)}.

\bibitem{Baumann:2009ds}
D.~Baumann,
``Inflation,''
\arxiv{0907.5424}{[hep-th]}.

\bibitem{Uzan:2015pfm}
J.~P.~Uzan,
``Inflation in the standard cosmological model,''
\journal{10.1016/j.crhy.2015.08.001}{Comptes Rendus Physique \textbf{16}, 875-890 (2015)}.


\bibitem{Mukhanov:1990me}
For a complete review on cosmological and inflationary perturbations see:\\
V.~F.~Mukhanov, H.~A.~Feldman and R.~H.~Brandenberger,
``Theory of cosmological perturbations. Part 1. Classical perturbations. Part 2. Quantum theory of perturbations. Part 3. Extensions,''
\journal{10.1016/0370-1573(92)90044-Z}{Phys. Rept. \textbf{215}, 203-333 (1992)}.

\bibitem{Riotto:2002yw}
For a comprehensive review on inflationary perturbations see:\\
A.~Riotto,
``Inflation and the theory of cosmological perturbations,''
\journal{}{ICTP Lect. Notes Ser. \textbf{14}, 317-413 (2003)}
\arxiv{hep-ph/0210162}{[hep-ph]}.

\bibitem{Guzzetti:2016mkm}
For a complete review on GW production during inflation see:\\
M.~C.~Guzzetti, N.~Bartolo, M.~Liguori and S.~Matarrese,
``Gravitational waves from inflation,''
\journal{10.1393/ncr/i2016-10127-1}{Riv. Nuovo Cim. \textbf{39}, no.9, 399-495 (2016)}
\arxiv{1605.01615}{[astro-ph.CO]}.


\bibitem{Steinhardt:1984jj}
P.~J.~Steinhardt and M.~S.~Turner,
``A Prescription for Successful New Inflation,''
\journal{10.1103/PhysRevD.29.2162}{Phys. Rev. D \textbf{29} (1984), 2162-2171}


\bibitem{Liddle:1994dx}
A.~R.~Liddle, P.~Parsons and J.~D.~Barrow,
``Formalizing the slow roll approximation in inflation,''
\journal{10.1103/PhysRevD.50.7222}{Phys. Rev. D \textbf{50}, 7222-7232 (1994)}
\arxiv{astro-ph/9408015}{[astro-ph]}.

\bibitem{Albrecht:1982mp}
A.~Albrecht, P.~J.~Steinhardt, M.~S.~Turner and F.~Wilczek,
``Reheating an Inflationary Universe,''
\journal{10.1103/PhysRevLett.48.1437}{Phys.\ Rev.\ Lett.\  {\bf 48} (1982) 1437}.


\bibitem{Abbott:1982hn}
L.~F.~Abbott, E.~Farhi and M.~B.~Wise,
``Particle Production in the New Inflationary Cosmology,''
\journal{10.1016/0370-2693(82)90867-X}{Phys.\ Lett.\  {\bf 117B} (1982) 29}.

\bibitem{Turner:1983he}
M.~S.~Turner,
``Coherent Scalar Field Oscillations in an Expanding Universe,''
\journal{doi:10.1103/PhysRevD.28.1243}{Phys.\ Rev.\ D {\bf 28} (1983) 1243}.

\bibitem{Shtanov:1993es}
Y.~Shtanov,
``Scalar-field dynamics and reheating of the universe in chaotic inflation scenario"
\journal{}{Ukr. Fiz. Zh., Vol. 38, No. 9, p. 1425 - 1434}.



\bibitem{Bassett:2005xm} 
B.~A.~Bassett, S.~Tsujikawa and D.~Wands,
``Inflation dynamics and reheating,''
\journal{10.1103/RevModPhys.78.537}{Rev.\ Mod.\ Phys.\  {\bf 78}, 537 (2006)}
\arxiv{astro-ph/0507632}{}.


\bibitem{Frolov:2010sz} 
A.~V.~Frolov,
``Non-linear Dynamics and Primordial Curvature Perturbations from Preheating,''
\journal{10.1088/0264-9381/27/12/124006}{Class.\ Quant.\ Grav.\  {\bf 27}, 124006 (2010)}
\arxiv{1004.3559}{[gr-qc]}.

\bibitem{Allahverdi:2010xz}
R.~Allahverdi, R.~Brandenberger, F.~Y.~Cyr-Racine and A.~Mazumdar,
``Reheating in Inflationary Cosmology: Theory and Applications,''
\journal{10.1146/annurev.nucl.012809.104511}{Ann. Rev. Nucl. Part. Sci. \textbf{60}, 27-51 (2010)}
\arxiv{1001.2600}{[hep-th]}.

\bibitem{Amin:2014eta} 
M.~A.~Amin, M.~P.~Hertzberg, D.~I.~Kaiser and J.~Karouby,
``Nonperturbative Dynamics Of Reheating After Inflation: A Review,''
\journal{10.1142/S0218271815300037}{Int.\ J.\ Mod.\ Phys.\ D {\bf 24}, 1530003 (2014)}
\arxiv{1410.3808}{[hep-ph]}.

\bibitem{Lozanov:2019jxc} 
K.~D.~Lozanov,
``Lectures on Reheating after Inflation,''
\arxiv{1907.04402}{[astro-ph.CO]}.



\bibitem{Sakharov:1967dj}
A.~D.~Sakharov,
``Violation of CP Invariance, C asymmetry, and baryon asymmetry of the universe,''
\journal{10.1070/PU1991v034n05ABEH002497}{Pisma Zh. Eksp. Teor. Fiz. \textbf{5}, 32-35 (1967)}

\bibitem{Ignatiev:1978uf}
A.~Y.~Ignatiev, N.~V.~Krasnikov, V.~A.~Kuzmin and A.~N.~Tavkhelidze,
``Universal CP Noninvariant Superweak Interaction and Baryon Asymmetry of the Universe,''
\journal{10.1016/0370-2693(78)90900-0}{Phys. Lett. B \textbf{76}, 436-438 (1978)}


\bibitem{Yoshimura:1978ex}
M.~Yoshimura,
``Unified Gauge Theories and the Baryon Number of the Universe,''
\journal{10.1103/PhysRevLett.41.281}{Phys. Rev. Lett. \textbf{41}, 281-284 (1978)}

\bibitem{Dimopoulos:1978kv}
S.~Dimopoulos and L.~Susskind,
``On the Baryon Number of the Universe,''
\journal{10.1103/PhysRevD.18.4500}{Phys. Rev. D \textbf{18}, 4500-4509 (1978)}



\bibitem{Yoshimura:1979gy}
M.~Yoshimura,
``Origin of Cosmological Baryon Asymmetry,''
Phys. Lett. B \textbf{88}, 294-298 (1979)

\bibitem{Weinberg:1979bt}
S.~Weinberg,
``Cosmological Production of Baryons,''
\journal{10.1103/PhysRevLett.42.850}{Phys. Rev. Lett. \textbf{42}, 850-853 (1979)}


\bibitem{Kolb:1979qa}
E.~W.~Kolb and S.~Wolfram,
``Baryon Number Generation in the Early Universe,''
\journal{10.1016/0550-3213(82)90012-8}{Nucl. Phys. B \textbf{172} (1980), 224.}

\bibitem{Fry:1980ph}
J.~N.~Fry, K.~A.~Olive and M.~S.~Turner,
``Evolution of Cosmological Baryon Asymmetries,''
\journal{10.1103/PhysRevD.22.2953}{Phys. Rev. D \textbf{22}, 2953 (1980)}


\bibitem{Fry:1980bc}
J.~N.~Fry, K.~A.~Olive and M.~S.~Turner,
``Higgs Bosons and the Evolution of Baryon Asymmetries,''
\journal{10.1103/PhysRevD.22.2977}{Phys. Rev. D \textbf{22}, 2977 (1980)}

\bibitem{Harvey:1981yk}
J.~A.~Harvey, E.~W.~Kolb, D.~B.~Reiss and S.~Wolfram,
``Calculation of Cosmological Baryon Asymmetry in Grand Unified Gauge Models,''
\journal{10.1016/0550-3213(82)90375-3}{Nucl. Phys. B \textbf{201}, 16-100 (1982)}


\bibitem{Dolgov:1982th}
A.~D.~Dolgov and A.~D.~Linde,
``Baryon Asymmetry in Inflationary Universe,''
\journal{10.1016/0370-2693(82)90292-1}{Phys.\ Lett.\  {\bf 116B} (1982) 329}.


\bibitem{Kolb:1983ni}
E.~W.~Kolb and M.~S.~Turner,
``Grand Unified Theories and the Origin of the Baryon Asymmetry,''
\journal{10.1146/annurev.ns.33.120183.003241}{Ann. Rev. Nucl. Part. Sci. \textbf{33}, 645-696 (1983)}

\bibitem{Affleck:1984fy}
I.~Affleck and M.~Dine,
``A New Mechanism for Baryogenesis,''
\journal{10.1016/0550-3213(85)90021-5}{Nucl. Phys. B \textbf{249}, 361-380 (1985)}

\bibitem{Linde:1985gh}
A.~D.~Linde,
``The New Mechanism of Baryogenesis and the Inflationary Universe,''
\journal{10.1016/0370-2693(85)91319-X}{Phys. Lett. B \textbf{160}, 243-248 (1985)}


\bibitem{Shaposhnikov:1987tw}
M.~E.~Shaposhnikov,
``Baryon Asymmetry of the Universe in Standard Electroweak Theory,''
\journal{10.1016/0550-3213(87)90127-1}{Nucl. Phys. B \textbf{287}, 757-775 (1987)}

\bibitem{Cohen:1987vi}
A.~G.~Cohen and D.~B.~Kaplan,
``Thermodynamic Generation of the Baryon Asymmetry,''
\journal{10.1016/0370-2693(87)91369-4}{Phys. Lett. B \textbf{199}, 251-258 (1987)}

\bibitem{Ellis:1987rw}
J.~R.~Ellis, K.~Enqvist, D.~V.~Nanopoulos and K.~A.~Olive,
``Inflationary Fluctuations, Entropy Generation and Baryogenesis,''
\journal{10.1016/0370-2693(87)90620-4}{Phys. Lett. B \textbf{191}, 343-348 (1987)}


\bibitem{Dolgov:1991fr}
A.~D.~Dolgov,
``NonGUT baryogenesis,''
Phys. Rept. \textbf{222} (1992), 309-386

\bibitem{Riotto:1998bt}
A.~Riotto,
``Theories of baryogenesis,''
[arXiv:hep-ph/9807454 [hep-ph]].
\arxiv{9807454}{[hep-ph]}

\bibitem{Riotto:1999yt}
A.~Riotto and M.~Trodden,
``Recent progress in baryogenesis,''
\journal{10.1146/annurev.nucl.49.1.35}{Ann. Rev. Nucl. Part. Sci. \textbf{49}, 35-75 (1999)}
\arxiv{9901362}{[hep-ph]}

\bibitem{Dine:2003ax}
M.~Dine and A.~Kusenko,
``The Origin of the matter - antimatter asymmetry,''
Rev. Mod. Phys. \textbf{76}, 1 (2003)
\arxiv{0303065}{[hep-ph]}

\bibitem{Cline:2006ts}
J.~M.~Cline,
``Baryogenesis,''
\arxiv{0609145}{[hep-ph]}

\bibitem{Shaposhnikov:2009zzb}
M.~Shaposhnikov,
``Baryogenesis,''
\journal{10.1088/1742-6596/171/1/012005}{J. Phys. Conf. Ser. \textbf{171}, 012005 (2009)}




\bibitem{Canetti:2012zc}
L.~Canetti, M.~Drewes and M.~Shaposhnikov,
``Matter and Antimatter in the Universe,''
\journal{10.1088/1367-2630/14/9/095012}{New J. Phys. \textbf{14}, 095012 (2012)}
\arxiv{1204.4186}{[hep-ph]}




\bibitem{Fukugita:1986hr}
M.~Fukugita and T.~Yanagida,
``Baryogenesis Without Grand Unification,''
\journal{10.1016/0370-2693(86)91126-3}{Phys.\ Lett.\ B {\bf 174} (1986) 45}

\bibitem{Luty:1992un}
M.~A.~Luty,
``Baryogenesis via leptogenesis,''
Phys. Rev. D \textbf{45}, 455-465 (1992);
\journal{10.1103/PhysRevD.45.455}{Phys. Rev. D \textbf{45}, 455-465 (1992)}

\bibitem{Covi:1996wh}
L.~Covi, E.~Roulet and F.~Vissani,
``CP violating decays in leptogenesis scenarios,''
\journal{10.1016/0370-2693(96)00817-9}{Phys. Lett. B \textbf{384}, 169-174 (1996)}
\arxiv{hep-ph/9605319}{[hep-ph]}

\bibitem{Plumacher:1996kc}
M.~Plumacher,
``Baryogenesis and lepton number violation,''
Z. Phys. C \textbf{74}, 549-559 (1997)
\journal{10.1007/s002880050418}{Z. Phys. C \textbf{74}, 549-559 (1997)}
\arxiv{9604229}{[hep-ph]}

\bibitem{Barbieri:1999ma}
R.~Barbieri, P.~Creminelli, A.~Strumia and N.~Tetradis,
``Baryogenesis through leptogenesis,''
Nucl. Phys. B \textbf{575}, 61-77 (2000)
\journal{10.1016/S0550-3213(00)00011-0}{Nucl. Phys. B \textbf{575}, 61-77 (2000)}
\arxiv{9911315}{[hep-ph]}

\bibitem{Giudice:2003jh}
G.~F.~Giudice, A.~Notari, M.~Raidal, A.~Riotto and A.~Strumia,
``Towards a complete theory of thermal leptogenesis in the SM and MSSM,''
\journal{10.1016/j.nuclphysb.2004.02.019}{Nucl. Phys. B \textbf{685}, 89-149 (2004)}
\arxiv{0310123}{[hep-ph]}




\bibitem{Lazarides:1990huy}
G.~Lazarides and Q.~Shafi,
``Origin of matter in the inflationary cosmology,''
\journal{10.1016/0370-2693(91)91090-I}{Phys. Lett. B \textbf{258}, 305-309 (1991)}


\bibitem{Murayama:1992ua}
H.~Murayama, H.~Suzuki, T.~Yanagida and J.~Yokoyama,
``Chaotic inflation and baryogenesis by right-handed sneutrinos,''
\journal{10.1103/PhysRevLett.70.1912}{Phys. Rev. Lett. \textbf{70}, 1912-1915 (1993)}


\bibitem{Campbell:1992hd}
B.~A.~Campbell, S.~Davidson and K.~A.~Olive,
``Inflation, neutrino baryogenesis, and (S)neutrino induced baryogenesis,''
\journal{10.1016/0550-3213(93)90619-Z}{Nucl. Phys. B \textbf{399}, 111-136 (1993)}
\arxiv{hep-ph/9302223}{[hep-ph]}


\bibitem{Kumekawa:1994gx}
K.~Kumekawa, T.~Moroi and T.~Yanagida,
``Flat potential for inflaton with a discrete R invariance in supergravity,''
[
\journal{10.1143/PTP.92.437}{Prog. Theor. Phys. \textbf{92}, 437-448 (1994)}
\arxiv{9405337}{ [hep-ph]}

\bibitem{Murayama:1993em}
H.~Murayama and T.~Yanagida,
``Leptogenesis in supersymmetric standard model with right-handed neutrino,''
\journal{10.1016/0370-2693(94)91164-9}{Phys. Lett. B \textbf{322}, 349-354 (1994)}
\arxiv{9310297}{[hep-ph]}

\bibitem{Giudice:1999fb}
G.~F.~Giudice, M.~Peloso, A.~Riotto and I.~Tkachev,
``Production of massive fermions at preheating and leptogenesis,''
\journal{10.1088/1126-6708/1999/08/014}{JHEP \textbf{08}, 014 (1999)}
\arxiv{9905242}{[hep-ph]}


\bibitem{Asaka:1999yd}
T.~Asaka, K.~Hamaguchi, M.~Kawasaki and T.~Yanagida,
``Leptogenesis in inflaton decay,''
\journal{10.1016/S0370-2693(99)01020-5}{Phys. Lett. B \textbf{464}, 12-18 (1999)}
\arxiv{9906366}{[hep-ph]}


\bibitem{Asaka:1999jb}
T.~Asaka, K.~Hamaguchi, M.~Kawasaki and T.~Yanagida,
``Leptogenesis in inflationary universe,''
\journal{10.1103/PhysRevD.61.083512}{Phys. Rev. D \textbf{61}, 083512 (2000)}
\arxiv{9907559}{[hep-ph]}

\bibitem{Zhang:2023oyo}
X.~Zhang,
``Towards a systematic study of non-thermal leptogenesis from inflaton decays,''
\journal{10.1007/JHEP05(2024)147}{JHEP \textbf{05}, 147 (2024)}
\arxiv{2311.05824}{[hep-ph]}



\bibitem{Buchmuller:2005eh}
W.~Buchmuller, R.~D.~Peccei and T.~Yanagida,
``Leptogenesis as the origin of matter,''
\journal{10.1146/annurev.nucl.55.090704.151558}{Ann. Rev. Nucl. Part. Sci. \textbf{55}, 311-355 (2005)}
\arxiv{hep-ph/0502169}{[hep-ph]}

\bibitem{Buchmuller:2004nz}
W.~Buchmuller, P.~Di Bari and M.~Plumacher,
``Leptogenesis for pedestrians,''
\journal{10.1016/j.aop.2004.02.003}{Annals Phys. \textbf{315}, 305-351 (2005)}
\arxiv{hep-ph/0401240}{[hep-ph]}

\bibitem{Davidson:2008bu} 
S.~Davidson, E.~Nardi and Y.~Nir,
``Leptogenesis,''
\journal{10.1016/j.physrep.2008.06.002}{Phys.Rept. {\bf 466}, 105 (2008).}
\arxiv{0802.2962}{[hep-ph]}

\bibitem{Fong:2012buy}
C.~S.~Fong, E.~Nardi and A.~Riotto,
``Leptogenesis in the Universe,''
\journal{10.1155/2012/158303}{Adv. High Energy Phys. \textbf{2012} (2012), 158303}
\arxiv{1301.3062}{[hep-ph]}

\bibitem{Blanchet:2012bk}
S.~Blanchet and P.~Di Bari,
``The minimal scenario of leptogenesis,''
\journal{10.1088/1367-2630/14/12/125012}{New J. Phys. \textbf{14}, 125012 (2012)}
\arxiv{1211.0512}{[hep-ph]}

\bibitem{Kuzmin:1985mm}
V.~A.~Kuzmin, V.~A.~Rubakov and M.~E.~Shaposhnikov,
``On the Anomalous Electroweak Baryon Number Nonconservation in the Early Universe,''
\journal{10.1016/0370-2693(85)91028-7}{Phys. Lett. B \textbf{155} (1985), 36.}


\bibitem{Arnold:1987mh}
P.~B.~Arnold and L.~D.~McLerran,
``Sphalerons, Small Fluctuations and Baryon Number Violation in Electroweak Theory,''
\journal{10.1103/PhysRevD.36.581}{Phys. Rev. D \textbf{36}, 581 (1987)}


\bibitem{Arnold:1987zg}
P.~B.~Arnold and L.~D.~McLerran,
``The Sphaleron Strikes Back,''
\journal{10.1103/PhysRevD.37.1020}{Phys. Rev. D \textbf{37}, 1020 (1988)}

\bibitem{Fukugita:1990gb}
M.~Fukugita and T.~Yanagida,
``Sphaleron Induced Baryon Number Nonconservation and a Constraint on Majorana Neutrino Masses,''
\journal{10.1103/PhysRevD.42.1285}{Phys. Rev. D \textbf{42}, 1285-1286 (1990)}



\bibitem{Fields:2019pfx}
B.~D.~Fields, K.~A.~Olive, T.~H.~Yeh and C.~Young,
``Big-Bang Nucleosynthesis after Planck,''
\journal{10.1088/1475-7516/2020/03/010}{JCAP \textbf{03}, 010 (2020)}
\arxiv{1912.01132}{[astro-ph.CO]}


\bibitem{Cooke:2017cwo}
R.~J.~Cooke, M.~Pettini and C.~C.~Steidel,
``One Percent Determination of the Primordial Deuterium Abundance,''
\journal{10.3847/1538-4357/aaab53}{Astrophys. J. \textbf{855}, no.2, 102 (2018)}
\arxiv{1710.11129}{[astro-ph.CO]}


\bibitem{ParticleDataGroup:2024cfk}
S.~Navas \textit{et al.} [Particle Data Group],
``Review of particle physics,''
\journal{10.1103/PhysRevD.110.030001}{Phys. Rev. D \textbf{110}, no.3, 030001 (2024)}





\bibitem{DESI:2024mwx}
A.~G.~Adame \textit{et al.} [DESI],
``DESI 2024 VI: cosmological constraints from the measurements of baryon acoustic oscillations,''
\journal{10.1088/1475-7516/2025/02/021}{JCAP \textbf{02}, 021 (2025)}
\arxiv{2404.03002}{[astro-ph.CO]}


\bibitem{Allali:2024aiv}
I.~J.~Allali and A.~Notari,
``Neutrino mass bounds from DESI 2024 are relaxed by Planck PR4 and cosmological supernovae,''
\journal{10.1088/1475-7516/2024/12/020}{JCAP \textbf{12}, 020 (2024)}
\arxiv{2406.14554}{[astro-ph.CO]}

\bibitem{Esteban:2024eli}
I.~Esteban, M.~C.~Gonzalez-Garcia, M.~Maltoni, I.~Martinez-Soler, J.~P.~Pinheiro and T.~Schwetz,
``NuFit-6.0: updated global analysis of three-flavor neutrino oscillations,''
\journal{10.1007/JHEP12(2024)216}{JHEP \textbf{12}, 216 (2024)}
\arxiv{2410.05380}{[hep-ph]}

\bibitem{Capozzi:2025wyn}
F.~Capozzi, W.~Giar{\`e}, E.~Lisi, A.~Marrone, A.~Melchiorri and A.~Palazzo,
``Neutrino masses and mixing: Entering the era of subpercent precision,''
\journal{10.1103/PhysRevD.111.093006}{Phys. Rev. D \textbf{111}, no.9, 093006 (2025)}
\arxiv{2503.07752}{[hep-ph]}

\bibitem{Capozzi:2025ovi}
F.~Capozzi, E.~Lisi, F.~Marcone, A.~Marrone and A.~Palazzo,
``Updated bounds on the (1,2) neutrino oscillation parameters after first JUNO results,''
\arxiv{2511.21650}{[hep-ph]}



\bibitem{KATRIN:2024cdt}
M.~Aker \textit{et al.} [KATRIN],
``Direct neutrino-mass measurement based on 259 days of KATRIN data,''
\journal{10.1126/science.adq9592}{Science \textbf{388}, no.6743, adq9592 (2025)}
\arxiv{2406.13516}{[nucl-ex]}


\bibitem{Minkowski:1977sc}
P.~Minkowski,
``$\mu \to e\gamma$ at a Rate of One Out of $10^{9}$ Muon Decays?,''
\journal{10.1016/0370-2693(77)90435-X}{Phys. Lett. B \textbf{67} (1977), 421-428}

\bibitem{Frampton:2002qc}
P.~H.~Frampton, S.~L.~Glashow and T.~Yanagida,
``Cosmological sign of neutrino CP violation,''
\journal{10.1016/S0370-2693(02)02853-8}{Phys. Lett. B \textbf{548} (2002), 119-121}
\arxiv{hep-ph/0208157}{[hep-ph]}


\bibitem{Raidal:2002xf}
M.~Raidal and A.~Strumia,
``Predictions of the most minimal seesaw model,''
Phys. Lett. B \textbf{553} (2003), 72-78
\journal{10.1016/S0370-2693(02)03124-6}{Phys. Lett. B \textbf{553} (2003), 72-78}
\arxiv{hep-ph/0210021}{[hep-ph]}




\bibitem{Yanagida:1980xy}
T.~Yanagida,
``Horizontal Symmetry and Masses of Neutrinos,''
Prog. Theor. Phys. \textbf{64} (1980), 1103;
\journal{10.1143/PTP.64.1103}{Prog. Theor. Phys. \textbf{64} (1980), 1103}

\bibitem{Glashow:1979nm}
S.~L.~Glashow,
``The Future of Elementary Particle Physics,''
NATO Sci. Ser. B \textbf{61} (1980), 687;
\journal{10.1007/978-1-4684-7197-7{\_}15}{NATO Sci. Ser. B \textbf{61} (1980), 687}

\bibitem{Gell-Mann:1979vob}
M.~Gell-Mann, P.~Ramond and R.~Slansky,
``Complex Spinors and Unified Theories,''
\journal{}{Conf. Proc. C \textbf{790927} (1979), 315-321}
\arxiv{1306.4669}{[hep-th]}




\bibitem{Pontecorvo:1957qd}
B.~Pontecorvo,
``Inverse Beta Processes and Nonconservation of Lepton Charge,''
Sov. Phys. JETP \textbf{7} (1958), 172-173;
\journal{}{Sov. Phys. JETP \textbf{7} (1958), 172-173}

\bibitem{Pontecorvo:1957cp}
B.~Pontecorvo,
``Mesonium and Antimesonium,''
Sov. Phys. JETP \textbf{6} (1958), 429-431;
\journal{}{Sov. Phys. JETP \textbf{6} (1958), 429-431}

\bibitem{Maki:1962mu}
Z.~Maki, M.~Nakagawa and S.~Sakata,
``Remarks on the unified model of elementary particles,''
\journal{10.1143/PTP.28.870}{Prog. Theor. Phys. \textbf{28} (1962), 870-880.}


\bibitem{Casas:2001sr}
J.~A.~Casas and A.~Ibarra,
``Oscillating neutrinos and $\mu \to e, \gamma$,''
\journal{10.1016/S0550-3213(01)00475-8}{Nucl. Phys. B \textbf{618}, 171-204 (2001)}
\arxiv{hep-ph/0103065}{[hep-ph]}



\bibitem{Davidson:2002qv}
S.~Davidson and A.~Ibarra,
``A Lower bound on the right-handed neutrino mass from leptogenesis,''
\journal{10.1016/S0370-2693(02)01735-5}{Phys. Lett. B \textbf{535}, 25-32 (2002)}
\arxiv{hep-ph/0202239}{[hep-ph]}


\bibitem{Weinberg:1982id}
S.~Weinberg,
``Does Gravitation Resolve the Ambiguity Among Supersymmetry Vacua?,''
\journal{10.1103/PhysRevLett.48.1776}{Phys. Rev. Lett. \textbf{48} (1982), 1776-1779.}

\bibitem{Ellis:1982yb}
J.~R.~Ellis, A.~D.~Linde and D.~V.~Nanopoulos,
``Inflation Can Save the Gravitino,''
\journal{10.1016/0370-2693(82)90601-3}{Phys. Lett. B \textbf{118}, 59-64 (1982)}

\bibitem{Nanopoulos:1983up}
D.~V.~Nanopoulos, K.~A.~Olive and M.~Srednicki,
``After Primordial Inflation,''
\journal{10.1016/0370-2693(83)91624-6}{Phys. Lett. B \textbf{127} (1983), 30-34.}

\bibitem{Ellis:1984eq}
J.~R.~Ellis, J.~E.~Kim and D.~V.~Nanopoulos,
``Cosmological Gravitino Regeneration and Decay,''
\journal{10.1016/0370-2693(84)90334-4}{Phys. Lett. B \textbf{145} (1984), 181-186.}

\bibitem{Khlopov:1984pf}
M.~Y.~Khlopov and A.~D.~Linde,
``Is It Easy to Save the Gravitino?,''
\journal{10.1016/0370-2693(84)91656-3}{Phys. Lett. B \textbf{138} (1984), 265-268.}

\bibitem{Kawasaki:1994af}
M.~Kawasaki and T.~Moroi,
``Gravitino production in the inflationary universe and the effects on big bang nucleosynthesis,''
\journal{10.1143/PTP.93.879}{Prog. Theor. Phys. \textbf{93} (1995), 879-900.}
\arxiv{hep-ph/9403364}{[hep-ph]}


\bibitem{Giudice:2008gu}
G.~F.~Giudice, L.~Mether, A.~Riotto and F.~Riva,
``Supersymmetric Leptogenesis and the Gravitino Bound,''
Phys. Lett. B \textbf{664} (2008), 21-24.
\journal{10.1016/j.physletb.2008.05.009}{Phys. Lett. B \textbf{664} (2008), 21-24.}
\arxiv{0804.0166}{[hep-ph]}


\bibitem{Kawasaki:2004qu}
M.~Kawasaki, K.~Kohri and T.~Moroi,
``Big-Bang nucleosynthesis and hadronic decay of long-lived massive particles,''
\arxiv{0408426}{[astro-ph]}
\journal{10.1103/PhysRevD.71.083502}{Phys. Rev. D \textbf{71}, 083502 (2005)}

\bibitem{Kawasaki:2008qe}
M.~Kawasaki, K.~Kohri, T.~Moroi and A.~Yotsuyanagi,
``Big-Bang Nucleosynthesis and Gravitino,''
\arxiv{0804.3745}{[hep-ph]}
\journal{10.1103/PhysRevD.78.065011}{Phys. Rev. D \textbf{78}, 065011 (2008)}



\bibitem{Kawasaki:2017bqm}
M.~Kawasaki, K.~Kohri, T.~Moroi and Y.~Takaesu,
``Revisiting Big-Bang Nucleosynthesis Constraints on Long-Lived Decaying Particles,''
\arxiv{1709.01211}{[hep-ph]}
\journal{10.1103/PhysRevD.97.023502}{Phys. Rev. D \textbf{97}, no.2, 023502 (2018)}

\bibitem{Khlopov:2025pub}
For a review, see \textit{e.g.} M.~Y.~Khlopov,
``Cosmoparticle Physics,''
World Scientific, 2025,
\journal{10.1142/11875}{World Scientific, 2025}




\bibitem{Pradisi:2022nmh}
G.~Pradisi and A.~Salvio,
``(In)equivalence of metric-affine and metric effective field theories,''
\journal{10.1140/epjc/s10052-022-10825-9}{Eur. Phys. J. C \textbf{82}, no.9, 840 (2022)}
\arxiv{2206.15041}{[hep-th]}.

\bibitem{DiMarco:2023ncs}
A.~Di Marco, E.~Orazi and G.~Pradisi,
``Einstein{\textendash}Cartan pseudoscalaron inflation,''
\journal{10.1140/epjc/s10052-024-12482-6}{Eur. Phys. J. C \textbf{84}  no.2, 146 (2024)}
\arxiv{2309.11345}{[hep-th]}.

\bibitem{Salvio:2022suk}
A.~Salvio,
``Inflating and reheating the Universe with an independent affine connection,''
\journal{10.1103/PhysRevD.106.103510}{Phys. Rev. D \textbf{106}, no.10, 103510 (2022)}   
\arxiv{2207.08830}{[hep-ph]}.

\bibitem{Salvio:2025izr}
A.~Salvio,
``Independent connection in action during inflation,''
\journal{112/Phys.Rev.D.112.L061301}{Phys. Rev. D \textbf{112} (2025) no.6, L061301} 
[arXiv:2504.10488 [hep-ph]].


\bibitem{Hehl:1976kj}
F.~W.~Hehl, P.~Von Der Heyde, G.~D.~Kerlick and J.~M.~Nester,
``General Relativity with Spin and Torsion: Foundations and Prospects,''
\journal{10.1103/RevModPhys.48.393}{Rev. Mod. Phys. \textbf{48} (1976), 393-416}.

\bibitem{Shapiro:2001rz}
I.~L.~Shapiro,
``Physical aspects of the space-time torsion,''
\journal{10.1016/S0370-1573(01)00030-8}{Phys. Rept. \textbf{357}, 113 (2002)}
\arxiv{hep-th/0103093}{[hep-th]}.

\bibitem{Hammond:2002rm}
R.~T.~Hammond,
``Torsion gravity,''
\journal{10.1088/0034-4885/65/5/201}{Rept. Prog. Phys. \textbf{65}, 599-649 (2002)}.





\cite{Choudhury:2014hja}
\bibitem{Choudhury:2014hja}
S.~Choudhury, B.~K.~Pal, B.~Basu and P.~Bandyopadhyay,
``Quantum Gravity Effect in Torsion Driven Inflation and CP violation,''
\journal{10.1007/JHEP10(2015)194}{JHEP \textbf{10} (2015), 194}
\arxiv{1409.6036}{[hep-th]}


\bibitem{Shaposhnikov:2020frq}
M.~Shaposhnikov, A.~Shkerin, I.~Timiryasov and S.~Zell,
``Einstein-Cartan gravity, matter, and scale-invariant generalization~,''
\journal{10.1007/JHEP08(2021)162}{JHEP \textbf{10}, 177 (2020)}
\arxiv{2007.16158}{[hep-th]}.

\bibitem{Shaposhnikov:2020gts}
M.~Shaposhnikov, A.~Shkerin, I.~Timiryasov and S.~Zell,
``Higgs inflation in Einstein-Cartan gravity,''
\journal{10.1088/1475-7516/2021/10/E01}{JCAP \textbf{02}, 008 (2021)}
\arxiv{2007.14978}{[hep-ph]}.

\bibitem{Shaposhnikov:2020aen}
M.~Shaposhnikov, A.~Shkerin, I.~Timiryasov and S.~Zell,
``Einstein-Cartan Portal to Dark Matter,''
\journal{10.1103/PhysRevLett.127.169901}{Phys. Rev. Lett. \textbf{126}, no.16, 161301 (2021)}
\arxiv{2008.11686}{[hep-ph]}.



\bibitem{Karananas:2021zkl}
G.~K.~Karananas, M.~Shaposhnikov, A.~Shkerin and S.~Zell,
``Matter matters in Einstein-Cartan gravity,''
\journal{10.1103/PhysRevD.104.064036}{Phys. Rev. D \textbf{104} (2021) no.6, 064036}
\arxiv{2106.13811}{[hep-th]}.

\bibitem{Desai:2015haa}
S.~Desai and N.~J.~Pop\l{}awski,
``Non-parametric reconstruction of an inflaton potential from Einstein\textendash{}Cartan\textendash{}Sciama\textendash{}Kibble gravity with particle production,''
\journal{10.1016/j.physletb.2016.02.014}{Phys. Lett. B \textbf{755} (2016), 183-189}
\arxiv{1510.08834}{[astro-ph.CO]}

\bibitem{Piani:2022gon}
M.~Piani and J.~Rubio,
``Higgs-Dilaton inflation in Einstein-Cartan gravity,''
\journal{10.1088/1475-7516/2022/05/009}{JCAP \textbf{05} (2022) no.05, 009}
\arxiv{2202.04665}{[gr-qc]}



\bibitem{He:2023vlj}
M.~He, K.~Kamada and K.~Mukaida,
``Quantum corrections to Higgs inflation in Einstein-Cartan gravity,''
\journal{10.1007/JHEP01(2024)014}{JHEP \textbf{01}, 014 (2024)}
\arxiv{2308.14398}{[hep-ph]}




\bibitem{Olmo:2022ops}
G.~J.~Olmo, E.~Orazi and G.~Pradisi,
``Conformal metric-affine gravities,''
\journal{10.1088/1475-7516/2022/10/057}{JCAP \textbf{10} (2022), 057}
\arxiv{2207.12597}{[hep-th]}.




\bibitem{Hojman:1980kv}
R.~Hojman, C.~Mukku and W.~A.~Sayed,
``Parity violation in metric torsion theories of gravitation,''
\journal{10.1103/PhysRevD.22.1915}{Phys. Rev. D \textbf{22} (1980), 1915-1921}.

\bibitem{Nelson:1980ph}
P.~C.~Nelson,
``Gravity With Propagating Pseudoscalar Torsion,''
\journal{10.1016/0375-9601(80)90348-5}{Phys. Lett. A \textbf{79} (1980), 285}.

\bibitem{Holst:1995pc}
S.~Holst,
``Barbero's Hamiltonian derived from a generalized Hilbert-Palatini action,''
\journal{10.1103/PhysRevD.53.5966}{Phys. Rev. D \textbf{53}, 5966-5969 (1996)}
\arxiv{gr-qc/9511026}{[gr-qc]}.







\bibitem{Langvik:2020nrs}
M.~L\r{a}ngvik, J.~M.~Ojanper\"a, S.~Raatikainen and S.~Rasanen,
``Higgs inflation with the Holst and the Nieh\textendash{}Yan term,''
\journal{10.1103/PhysRevD.103.083514}{Phys. Rev. D \textbf{103}, no.8, 083514 (2021)}
\arxiv{2007.12595}{[astro-ph.CO]}.

\bibitem{Karananas:2021gco}
G.~K.~Karananas, M.~Shaposhnikov, A.~Shkerin and S.~Zell,
``Scale and Weyl invariance in Einstein-Cartan gravity,''
\journal{10.1103/PhysRevD.104.124014}{Phys. Rev. D \textbf{104} (2021) no.12, 124014}
\arxiv{2108.05897}{[hep-th]}.


\bibitem{Gialamas:2022xtt}
I.~D.~Gialamas and K.~Tamvakis,
``Inflation in metric-affine quadratic gravity,''
\journal{10.1088/1475-7516/2023/03/042}{JCAP \textbf{03} (2023), 042}
\arxiv{2212.0989}{[gr-qc]}

\bibitem{Gialamas:2024uar}
I.~D.~Gialamas and A.~Racioppi,
``Symmetry-breaking inflation in non-minimal metric-affine gravity,''
\journal{10.1088/1475-7516/2025/05/072}{JCAP \textbf{05}, 072 (2025)}
\arxiv{2412.17738}{[gr-qc]}

\bibitem{Gialamas:2024iyu}
I.~D.~Gialamas and K.~Tamvakis,
``Inflation in Weyl-invariant Einstein-Cartan gravity,''
\journal{10.1103/PhysRevD.111.044007}{Phys. Rev. D \textbf{111}, no.4, 044007 (2025)}
\arxiv{2410.16364}{[gr-qc]}


\bibitem{Gialamas:2026pjo}
I.~D.~Gialamas,
``Reheating in geometric Weyl-invariant Einstein-Cartan gravity,''
\journal{10.1103/4dz9-xb38}{Phys. Rev. D \textbf{113}, no.8, 084027 (2026)}
\arxiv{2602.00317}{[gr-qc]}

\bibitem{Racioppi:2025igu}
A.~Racioppi,
``Quasi-pole inflation in metric-affine gravity,''
\journal{10.1007/s10714-026-03548-9}{Gen. Rel. Grav. \textbf{58}, no.4, 44 (2026)}
\arxiv{2512.16815}{[gr-qc]}

\bibitem{Katsoulas:2025srh}
T.~Katsoulas and K.~Tamvakis,
``Inflationary assessment of $F(\mathcal{R},\tilde{\mathcal{R}})$ Einstein-Cartan models,''
\arxiv{2512.02847}{[gr-qc]}






\bibitem{BarberoG:1994eia}
J.~F.~Barbero G.,
``Real Ashtekar variables for Lorentzian signature space times,''
Phys. Rev. D \textbf{51} (1995), 5507-5510;
\journal{10.1103/PhysRevD.51.5507}{Phys. Rev. D \textbf{51} (1995), 5507-5510}
\arxiv{gr-qc/9410014}{[gr-qc]}.

\bibitem{Immirzi:1996di}
G.~Immirzi,
``Real and complex connections for canonical gravity,''
Class. Quant. Grav. \textbf{14} (1997), L177-L181.
\journal{10.1088/0264-9381/14/10/002}{Class. Quant. Grav. \textbf{14} (1997), L177-L181}
\arxiv{gr-qc/9612030}{[gr-qc]}.







\bibitem{Planck:2018jri}
Y.~Akrami \textit{et al.} [Planck],
``Planck 2018 results. X. Constraints on inflation,''
\journal{10.1051/0004-6361/201833887}{Astron. Astrophys. \textbf{641} (2020), A10}
\arxiv{1807.06211}{[astro-ph.CO]}.


\bibitem{AtacamaCosmologyTelescope:2025blo}
T.~Louis \textit{et al.} [Atacama Cosmology Telescope],
``The Atacama Cosmology Telescope: DR6 power spectra, likelihoods and {\ensuremath{\Lambda}}CDM parameters,''
\journal{10.1088/1475-7516/2025/11/062}{JCAP \textbf{11}, 062 (2025)}
\arxiv{2503.14452}{[astro-ph.CO]}

\bibitem{SPT-3G:2025bzu}
E.~Camphuis \textit{et al.} [SPT-3G],
``SPT-3G D1: CMB temperature and polarization power spectra and cosmology from 2019 and 2020 observations of the SPT-3G Main field,''
\arxiv{2506.20707}{[astro-ph.CO]}

\bibitem{Balkenhol:2025wms}
L.~Balkenhol, E.~Camphuis, F.~Finelli, K.~Benabed, F.~R.~Bouchet, J.~Carron, S.~Galli, E.~Hivon, A.~R.~Khalife and L.~Knox, \textit{et al.}
``Inflation at the End of 2025: Constraints on $r$ and $n_s$ Using the Latest CMB and BAO Data,''
\arxiv{2512.10613}{[astro-ph.CO]}

\bibitem{DiMarco:2024yzn}
A.~D.~Di Marco, E.~Orazi and G.~Pradisi,
``Introduction to the Number of e-Folds in Slow-Roll Inflation,''
Universe \textbf{10} (2024) no.7, 284.
\journal{10.3390/universe10070284}{Universe \textbf{10} (2024) no.7, 284.}
\arxiv{2408.01854}{[astro-ph.CO]}

\bibitem{Ferreira:2025lrd}
E.~G.~M.~Ferreira, E.~McDonough, L.~Balkenhol, R.~Kallosh, L.~Knox and A.~Linde,
``BAO-CMB tension and implications for inflation,''
\journal{10.1103/lq71-b84v}{Phys. Rev. D \textbf{113}, no.4, 043524 (2026)}
\arxiv{2507.12459}{[astro-ph.CO]}

\bibitem{Ye:2025ark}
G.~Ye and S.~J.~Lin,
``On the tension between DESI DR2 BAO and CMB,''
\arxiv{2505.02207}{[astro-ph.CO]}

\bibitem{Freedman:2012zz}
D.~Z.~Freedman and A.~Van Proeyen,
\emph{Supergravity},
\journal{10.1017/CBO9781139026833}{Cambridge Univ. Press, 2012}

\bibitem{Peskin:1995ev}
M.~E.~Peskin and D.~V.~Schroeder,
\emph{An Introduction to quantum field theory},
\journal{10.1201/9780429503559}{Addison-Wesley, 1995}

\bibitem{DiMarco:2021xzk}
A.~Di Marco and G.~Pradisi,
``Variable inflaton equation-of-state and reheating,''
\journal{10.1142/S0217751X21500950}{Int. J. Mod. Phys. A \textbf{36} no.15, 2150095 (2021)}
\arxiv{2102.00326}{[gr-qc]}.

\bibitem{Abramowitz:1965zz}
M.~Abramowitz and I.~A.~Stegun,
\emph{Handbook of Mathematical Functions: With Formulas, Graphs, and Mathematical Tables},
\journal{}{Dover Publications, 1965}








\end{thebibliography}
\end{document}